\PassOptionsToPackage{dvipsnames,table}{xcolor}

\documentclass[journal=jacsat,manuscript=article]{achemso}

\mciteErrorOnUnknownfalse
\usepackage{graphicx}
\usepackage{amsmath,amssymb}
\usepackage{booktabs}
\usepackage{array}
\usepackage{listings}
\usepackage[version=3]{mhchem}
\usepackage{longtable}

\usepackage{pifont}       

\newcommand{\cmark}{\textcolor{ForestGreen}{\Large\ding{51}}}   
\newcommand{\xmark}{\textcolor{Red}{\Large\ding{55}}}           

\usepackage[most]{tcolorbox}
\tcbset{
  precursor tile/.style={
    colback=blue!5!white,
    colframe=blue!75!black,
    boxrule=0.5pt,
    arc=2pt,
    width=5cm,
    halign=center,
    valign=center,
    fonttitle=\bfseries,
    boxsep=4pt,
  }
}

\usepackage{colortbl}                  

\definecolor{deepforest}{HTML}{76B041}
\definecolor{deepfire}{HTML}{B22222}
\definecolor{validpurple}{HTML}{6C48FF}   
\definecolor{invalidyellow}{HTML}{FFB200} 

\DeclareRobustCommand{\impGrad}[1]{%
  \begingroup
    \edef\val{#1}%
    \ifdim\val pt>10pt \cellcolor{deepforest!100!white}%
    \else\ifdim\val pt>7.5pt \cellcolor{deepforest!80!white}%
    \else\ifdim\val pt>5pt \cellcolor{deepforest!60!white}%
    \else\ifdim\val pt>2.5pt \cellcolor{deepforest!40!white}%
    \else\ifdim\val pt>0pt \cellcolor{deepforest!20!white}\fi\fi\fi\fi\fi
    \ifdim\val pt<0pt
      \ifdim\val pt>-2.5pt \cellcolor{deepfire!20!white}%
      \else\ifdim\val pt>-5pt \cellcolor{deepfire!40!white}%
      \else\ifdim\val pt>-7.5pt \cellcolor{deepfire!60!white}%
      \else\ifdim\val pt>-10pt \cellcolor{deepfire!80!white}%
      \else \cellcolor{deepfire!100!white}\fi\fi\fi\fi
    \fi
    \val\%%
  \endgroup
}

\usepackage{adjustbox}
\usepackage{booktabs,makecell,adjustbox}
\usepackage{chemformula}

\author{Thorben Prein}
\affiliation[TUM]{Technische Universität München, Garching b. München, Germany} 
\alsoaffiliation[MDSI]{Munich Data Science Institute, Technische Universität München, Munich, Germany} 
\alsoaffiliation[TUMintER]{TUMint. Energy Research GmbH, Munich, Germany} 

\author{Elton Pan}
\affiliation[MIT]{Department of Materials Science and Engineering, Massachusetts Institute of Technology, Cambridge, MA, USA} 
\author{Janik Jehkul}     
\affiliation[TUM]{Technische Universität München, Garching b. München, Germany}

\author{Steffen Weinmann}     
\affiliation[TUM]{Technische Universität München, Garching b. München, Germany}

\author{Elsa A. Olivetti}
\affiliation[MIT]{Department of Materials Science and Engineering, Massachusetts Institute of Technology, Cambridge, MA, USA} 

\author{Jennifer L. M. Rupp}
\email{jrupp@tum.de}
\affiliation[TUM]{Technische Universität München, Garching b. München, Germany} 
\alsoaffiliation[TUMintER]{TUMint. Energy Research GmbH, Munich, Germany}

\title[An \textsf{achemso} demo]
  {Language Models Enable Data-Augmented Synthesis Planning for Inorganic Materials}

\abbreviations{IR,NMR,UV}
\usepackage[colorlinks=true,allcolors=validpurple]{hyperref}
\begin{document}



\begin{abstract}
Inorganic synthesis planning currently relies primarily on heuristic approaches or machine‐learning models trained on limited datasets, which constrains its generality. We demonstrate that language models, without task-specific fine-tuning, can recall synthesis conditions. Off-the-shelf models, such as \textit{GPT-4.1}, \textit{Gemini 2.0 Flash} and \textit{Llama 4 Maverick}, achieve a Top-1 precursor-prediction accuracy of up to 53.8\,\% and a Top-5 performance of 66.1\,\% on a held-out set of 1,000 reactions. They also predict calcination and sintering temperatures with mean absolute errors below 126\,\textdegree C, matching specialized regression methods. Ensembling these language models further enhances predictive accuracy and reduces inference cost per prediction by up to 70\%. We subsequently employ language models to generate 28,548 synthetic reaction recipes, which we combine with literature-mined examples to pretrain a transformer-based model, \texttt{SyntMTE}. After fine‐tuning on the combined dataset, \texttt{SyntMTE} reduces mean‐absolute error in sintering temperature prediction to 73\,\textdegree C and in calcination temperature to 98\,\textdegree C. This strategy improves models by up to 8.7\,\% compared with baselines trained exclusively on experimental data. Finally, in a case study on \ce{Li7La3Zr2O12} solid-state electrolytes, we demonstrate that \texttt{SyntMTE} reproduces the experimentally observed dopant-dependent sintering trends. Our hybrid workflow enables scalable, data-efficient inorganic synthesis planning.
\end{abstract}


\begin{figure}[htbp]
    \centering
    \includegraphics[width=0.8\textwidth]{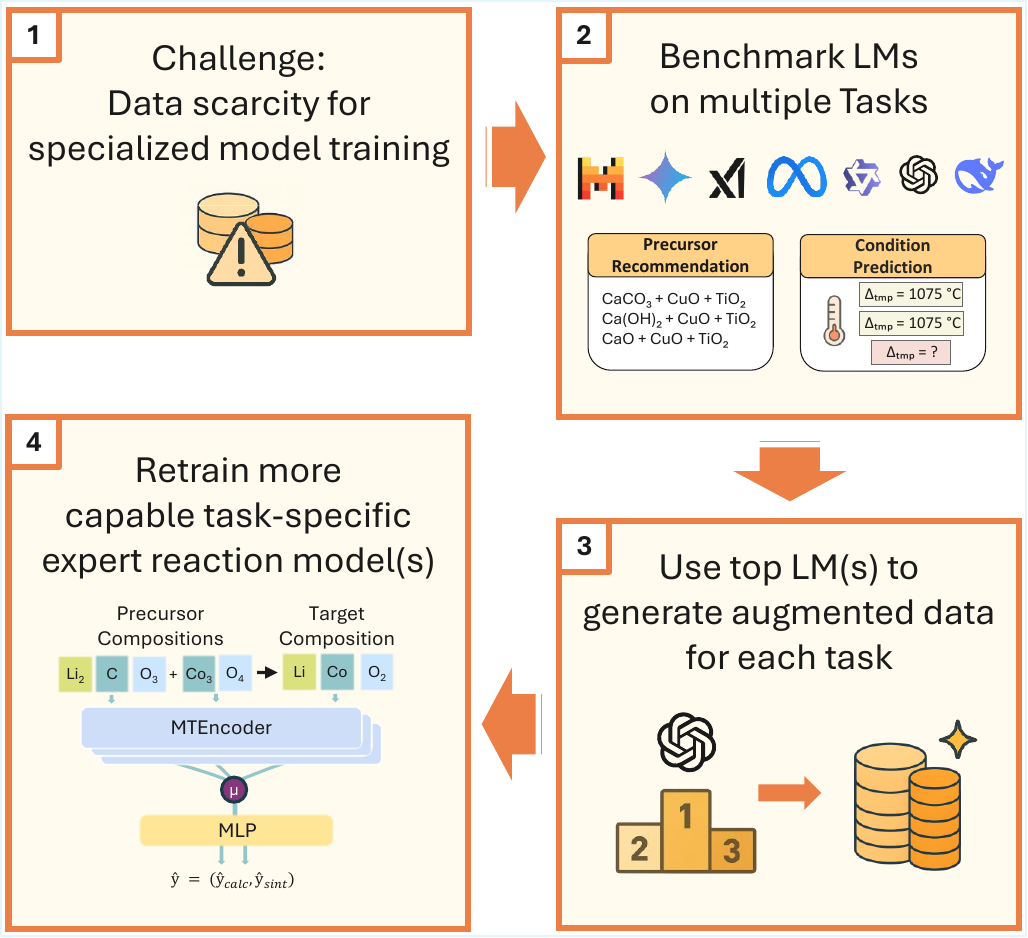} 
    \caption{Graphical Abstract.}
\end{figure}

\section{Introduction}

The discovery and design of advanced materials underpin progress in energy conversion and storage, information technology and medicine \cite{zhou2014interface, de2013high, whitesides2003right, goodenough2011challenges, cui2003high}. 
Recent advances in machine learning (ML) accelerated simulations have driven a rapid increase in candidate materials computationally predicted, now exceeding $4\times10^{6}$ stable structures \cite{sun2017thermodynamic,karan2025ion}. As a result, the synthesis of these candidates has become the principal bottleneck in the materials discovery pipeline \cite{mcdermott2023assessing,karpovich2023interpretable,abed2024open,malik2021predicting, szymanski2023autonomous}. Although density functional theory provides valuable thermodynamic insight, it remains challenging to accurately predict kinetics, diffusion, or phase-transformation pathways, leaving synthesis largely a trial-and-error process \cite{szczypinski2021can,woods2022role,chen2024navigating,kononova2019text,karpovich2023interpretable}. 

Accordingly, researchers have adopted ML methods to extract synthesis protocols from the scientific literature and predict feasible reaction pathways for novel structures \cite{olivetti2020data, kononova2021opportunities, sun2025critical}. Foundational studies by Kononova et al. \cite{kononova2019text}, E. Kim et al. \cite{kim2017materials}, and Huo et al. \cite{huo2019semi} curated comprehensive synthesis databases, thereby enabling ML-based approaches to inorganic synthesis planning. Recent efforts have focused on ML methods, leading to two principal tasks: (i) precursor recommendation, i.e., identifying suitable reagent combinations; and (ii) synthesis-condition prediction, i.e., determining optimal reaction parameters. These tasks are then applied sequentially to propose a viable synthesis protocol for a target material.

\textbf{Precursor recommendation}
Most precursor recommendation methods focus on solid‐state synthesis. An example for LiCoO\textsubscript{2} is illustrated in Fig. \ref{tab:precursor_suggestion_gpt41}, where a model ranks suitable precursor combinations by their likelihood. E. Kim et al.\cite{kim2020inorganic} used an RNN with ELMo embeddings to extract over 50,000 synthesis actions and over 116,000 precursor mentions, then trained a paired conditional VAE to jointly model action sequences and precursor formulas, enabling plausible precursor suggestions for novel targets. S. Kim\cite{kim2022element} introduced an element‐wise retrosynthesis with 39 template classes; He et al.\cite{he2023precursor} developed a retrieval‐based model using attention to compare planned synthesis to historical routes, extended by Noh et al.\cite{noh2024retrieval} via an enthalpy‐aware second ranker. Prein et al.\cite{prein2025retro} improved generalization by embedding materials with a pretrained transformer using a pairwise ranker to assess precursor suitability for unseen compounds.

\textbf{Synthesis condition prediction}
After precursor selection, a second type of model predicts the isothermal‐hold temperatures and dwell times for both the calcination and sintering steps in solid-state synthesis routes. Huo et al.\cite{huo2022machine} applied linear and tree‐based regression on text‐mined features (e.g., melting points, formation energies), achieving a mean absolute error (MAE) of approximately 140 °C. Prein et al.\cite{prein2024reaction} developed a Reaction Graph Network using MTEncoder embeddings and graph‐attention layers, reducing the error to approximately 90 °C. Pan et al.\cite{pan2024chemically,pan2024zeosyn} framed condition prediction as a diffusion‐based generative task conditioned on target materials structure, cutting the Wasserstein distance by over 25 \% by capturing the one-to-many nature of structure-synthesis relationships.

Despite the preponderance of ML methods, models are still bottlenecked by data-centric concerns. Synthesis databases remain relatively small, only  rarely exceeding a few \(\times 10^3\) unique entries, leaving the majority of chemistries unrepresented (Fig. \ref{fig:gen_dataset_fig})\cite{karpovich2023interpretable, kononova2021opportunities}.
The limited size of existing datasets inhibits recovery of the true distribution of process parameters for most materials, thereby precluding robust mapping of a “synthesis window”, the region of temperature, and isothermal dwell-time combinations that yield the desired phase. This behavior is directly analogous to phase-stability regimes depicted in classical time–temperature–transformation (TTT) diagrams \cite{zhu2023time, rupp2010time}. 
Furthermore, automated text‑mining pipelines exacerbate data sparsity by introducing extraction errors, such as misassigned stoichiometries, omitted precursor references, and conflation of precursor and target species, particularly in complex multi‑step protocols. Consequently, machine‑learning models trained on these sparse, noisy datasets cannot confidently resolve the underlying "synthesis window”, leading to diminished predictive accuracy and poor generalization to novel materials.



\textbf{Language models in materials synthesis}
In contrast, language models (LMs) leverage orders of magnitude more unstructured chemical knowledge: implicit heuristics, phase‐diagram insights, and procedural narratives from their extensive pretraining corpora. They have demonstrated considerable utility across scientific disciplines \cite{mirza2025framework, gottweis2025towards, penades2025ai, cohrs2025large, taylor2022galactica}. Generative LMs have excelled in crystal‐structure generation: CrystaLLM\cite{antunes2024crystal} and Crystal-Text-LLM\cite{rubungo2024llm4mat} produce DFT-validated geometries, while FlowLLM\cite{sriram2024flowllm} refines LLM-generated structures using flow matching\cite{miller2024flowmm}. In synthesis planning, GPT variants fine-tuned for synthesizability and precursor prediction may match specialized models\cite{kim2024large}. ALDbench evaluates language models on atomic layer deposition recipe prediction and finds that GPT-4o scores reasonable 3.7 out of 5\cite{yanguas2025benchmarking}. Despite the promising first adaptations, LMs have never been systematically compared in solid state inorganic synthesis, the crucial domain for materials discovery \cite{szymanski2023autonomous}. Benchmarking LMs in this domain allows researchers to make informed decisions and informs judgment on LLM suggestion goodness. However, LM predictions alone cannot demonstrate real progress over existing workflows when there are no independent training and test sets and data leakage from earlier synthesis reports is likely. This motivates two key research questions:

\begin{enumerate}
  \item \textbf{Benchmarking LM performance:} How do state-of-the-art LMs perform on inorganic solid-state synthesis tasks?
  \item \textbf{Data augmentation via LM generation:} Can LMs generate high-quality synthetic synthesis recipes to expand and enrich literature-mined databases, thereby mitigating data scarcity and sparsity?
\end{enumerate}



In this work, we explore LMs for inorganic synthesis planning and use them to expand literature‐mined solid‐state databases, finding that these synthetic entries boost prediction accuracy.
Finally, leveraging our synthesis‐planning framework, we reconstruct synthesis trends for doped \ce{Li7La3Zr2O12} (LLZO) compounds, a functional ceramic whose cubic phase is challenging to stabilize and therefore requires careful selection of dopants and sintering steps \cite{kim2021solid, Weinmann2025Sustainable, Balaish2025Emerging}. Overall, our contributions are as follows:

\begin{itemize}
  \item We demonstrate strong performance of LMs in recalling previously reported synthesis trends. Moreover, we demonstrate that an ensemble of LMs surpasses individual models in precursor suggestion and processing-condition prediction, and can even accurately reconstruct synthesis condition distributions from the literature.
  \item We curate a synthetic dataset of 28,548 complete LM-generated solid-state synthesis recipes, which signifies a 616$\%$ increase in complete entries over existing solid-state synthesis datasets \cite{kononova2019text}.
\item Leveraging both literature-mined and synthetic data, we develop a transformer-based model (\texttt{SyntMTE}) to regress synthesis conditions for inorganic synthesis. This model outperforms baseline methods such as CrabNet, composition-based neural networks, and decision-tree regressors, while all models exhibit significant improvements up to \>8.7\% when initially trained on the synthetic dataset.
 
\end{itemize}

\begin{table}[ht]
  \centering
  \caption{Top 10 Predicted Precursor Sets for LiCoO\textsubscript{2} using GPT-4.1}
  \label{tab:precursor_suggestion_gpt41}
  \renewcommand{\arraystretch}{1.2}
  \setlength{\tabcolsep}{6pt}
  \begin{tabular}{c  c  c  c}
    \toprule
    \textbf{Rank} & \textbf{Precursor A} & \textbf{Precursor B} & \textbf{Valid Path?} \\
    \midrule
    1  & \begin{tcolorbox}[precursor tile]Li$_2$CO$_3$\end{tcolorbox}
       & \begin{tcolorbox}[precursor tile]Co$_3$O$_4$\end{tcolorbox}
       & \cmark \\
    2  & \begin{tcolorbox}[precursor tile]Li$_2$CO$_3$\end{tcolorbox}
       & \begin{tcolorbox}[precursor tile]CoO\end{tcolorbox}
       & \cmark \\
    3  & \begin{tcolorbox}[precursor tile]Li$_2$CO$_3$\end{tcolorbox}
       & \begin{tcolorbox}[precursor tile]Co$_2$O$_3$\end{tcolorbox}
       & \xmark \\
    4  & \begin{tcolorbox}[precursor tile]LiOH\end{tcolorbox}
       & \begin{tcolorbox}[precursor tile]Co$_3$O$_4$\end{tcolorbox}
       & \cmark \\
    5  & \begin{tcolorbox}[precursor tile]LiOH\end{tcolorbox}
       & \begin{tcolorbox}[precursor tile]CoO\end{tcolorbox}
       & \xmark \\
    6  & \begin{tcolorbox}[precursor tile]LiOH\end{tcolorbox}
       & \begin{tcolorbox}[precursor tile]Co$_2$O$_3$\end{tcolorbox}
       & \xmark \\
    7  & \begin{tcolorbox}[precursor tile]Li$_2$O\end{tcolorbox}
       & \begin{tcolorbox}[precursor tile]Co$_3$O$_4$\end{tcolorbox}
       & \cmark \\
    8  & \begin{tcolorbox}[precursor tile]Li$_2$O\end{tcolorbox}
       & \begin{tcolorbox}[precursor tile]CoO\end{tcolorbox}
       & \cmark \\
    9  & \begin{tcolorbox}[precursor tile]Li$_2$O\end{tcolorbox}
       & \begin{tcolorbox}[precursor tile]Co$_2$O$_3$\end{tcolorbox}
       & \xmark \\
   10  & \begin{tcolorbox}[precursor tile]LiNO$_3$\end{tcolorbox}
       & \begin{tcolorbox}[precursor tile]Co$_3$O$_4$\end{tcolorbox}
       & \cmark \\
    \bottomrule
  \end{tabular}
\end{table}

\section{Results and Discussion}

\subsection{Benchmarking LMs in inorganic materials synthesis}

To assess LM capabilities in inorganic synthesis planning, we deploy state-of-the-art LMs on two synthesis tasks. We use a dataset derived from Kononova \cite{kononova2019text}, which contains roughly 10,000 unique precursor–target material combinations. For precursor recommendation, we downselect a 1000 datapoint test dataset in accordance with previous work \cite{prein2025retro}. We submit prompts to the LMs via OpenRouter without specifying the number of precursors, thereby requiring each model to infer the appropriate precursor count for each reaction. For synthesis condition prediction, a 1,000-entry dataset is curated by filtering for entries with both sintering and calcination temperatures present. For both tasks, we give the LMs 40 in-context examples from the held-out validation fraction of the dataset (Fig. \ref{fig:hyperparams-in-context}). For precursor prediction, we evaluate the exact match accuracy \cite{noh2024retrieval}. We decide to evaluate a total of 7 state-of-the-art LMs to cover a diverse umbrella of models, further details on the different LMs can be found in the Appendix.


\subsubsection{Precursor recommendation}

We evaluate the language models on the precursor prediction task and report Top-\(k\) exact-match accuracies. The exact-match accuracy can be considered a lower bound on performance, as the model is required to reproduce precisely the precursor set reported in the literature, whereas alternative, unreported synthesis routes may also exist \cite{pan2024chemically}.
Because practical precursor recommendation requires experimental validation of multiple candidate routes, Top-5 and Top-10 metrics are particularly informative, indicating whether a correct precursor set appears within the model’s top 5 or 10 suggestions. Table \ref{tab:model_performance_llm_precursor} shows that all models deliver competitive scores and cluster within a narrow performance band. Only Qwen~2.5~VL, scores lower by a clear margin, nevertheless its reported considerable parameter size. OpenAI~GPT-4.1 leads the Top-1 ranking at 53.8\% and retains good performance for high \(k\). The model is followed by Grok 3 mini and Llama~4 Maverick, as well as DeepSeek Chat~v3. In the Appendix, we compare the LM results with baseline methods from the literature. This comparison is only partly valid. Baseline models were trained on limited datasets. Furthermore, LMs may have benefited from data leakage during pretraining on test-set synthesis protocols. The best baseline reported in the literature achieves Top-5 and Top-10 accuracies of 73\% and 78\%, respectively, while individual language models attain close scores up to 66\% and 69\%. This example provides a compelling demonstration that state-of-the-art LMs, without any chemistry-specific training objectives, are able to recall high-quality chemistry knowledge through in-context learning only.

\begin{table*}[htb]
\centering
\caption{\textbf{Precursor recommendation performance.} Top‐$k$ exact‐match accuracies for individual language models and ensemble strategies on retrosynthesis precursor prediction. GPT-4.1 achieves the highest Top-1 accuracy, while min-rank ensembles boost performance at higher Top-$k$ thresholds. Notably, the ensemble of Llama~4 Maverick, DeepSeek Chat v3, and Gemini 2.0 Flash surpasses GPT-4.1 for relevant Top-5 and Top-10 settings with a 70\% reduction in inference cost.}
\label{tab:model_performance_llm_precursor}
\begingroup
\footnotesize
\resizebox{\linewidth}{!}{
\begin{tabular}{l c c c c}
\toprule
\textbf{Model} & \textbf{Top‐1} $\uparrow$ & \textbf{Top‐3} $\uparrow$ & \textbf{Top‐5} $\uparrow$ & \textbf{Top‐10} $\uparrow$\\
\midrule
{\raisebox{-0.3\height}{\includegraphics[height=1.5em]{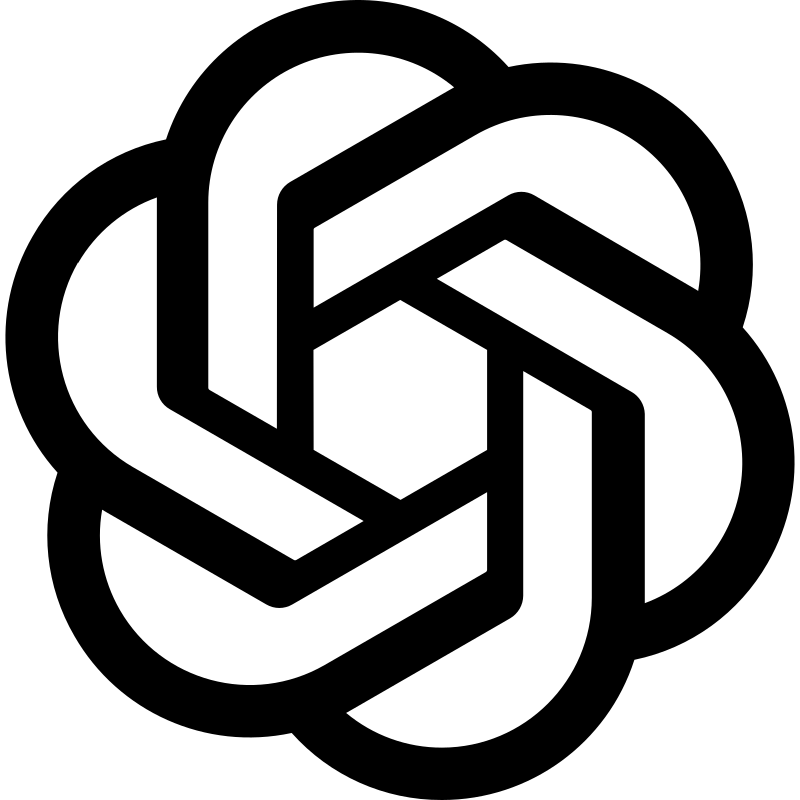}}\kern0.5em
\raisebox{-0.3\height}{\includegraphics[height=1.5em]{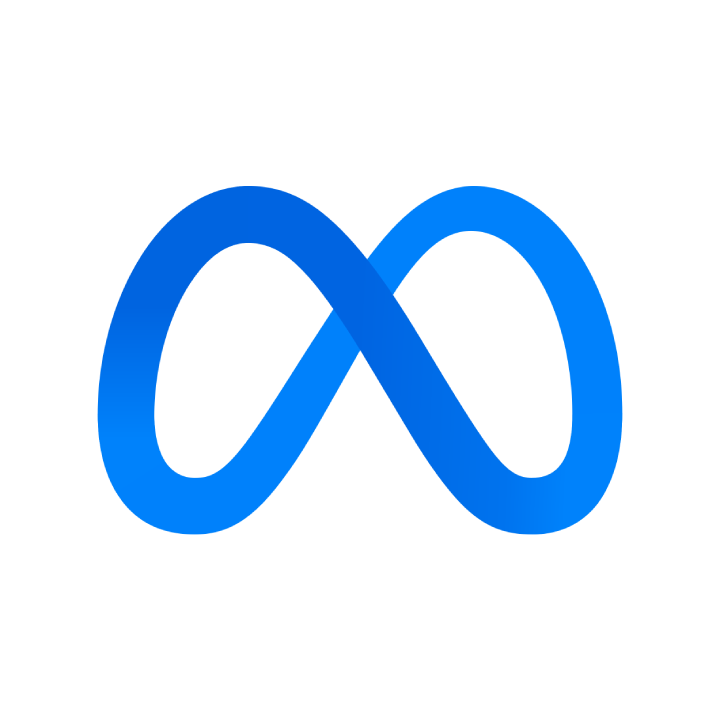}}\kern0.5em
\raisebox{-0.3\height}{\includegraphics[height=1.5em]{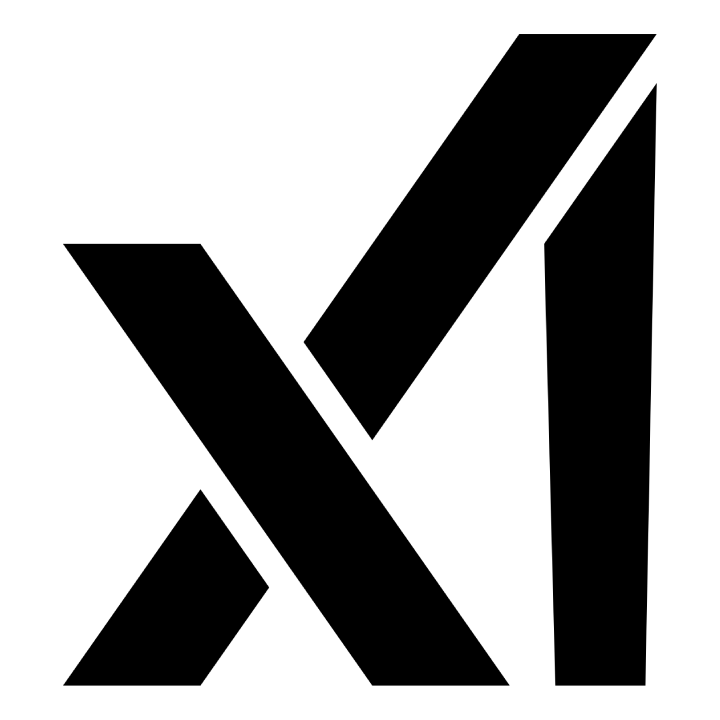}}}\quad
Ensemble Min-Rank
  & 52.3 
  & \textbf{65.8} 
  & \textbf{70.7}  
  & \textbf{74.3} \\ 
{\raisebox{-0.3\height}{\includegraphics[height=1.5em]{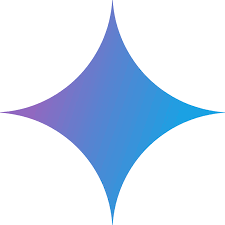}}\kern0.5em
 \raisebox{-0.3\height}{\includegraphics[height=1.5em]{figures/icons/meta.png}}\kern0.5em
 \raisebox{-0.3\height}{\includegraphics[height=1.5em]{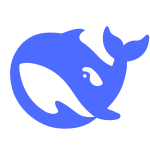}}}\quad
Ensemble Min-Rank
  & 51.8 
  & 63.1 
  & 67.4  
  & 71.9 \\ 
\midrule
\adjustbox{valign=m}{\includegraphics[height=1.5em]{figures/icons/openai.png}}\quad OpenAI GPT-4.1
  & \textbf{53.8}
  & \textbf{64.1}
  & \underline{66.1}
  & 68.7 \\[0.5ex]

\adjustbox{valign=m}{\includegraphics[height=1.5em]{figures/icons/xai.png}}\quad Grok 3 Mini Beta
  & 52.2
  & \underline{63.2}
  & \textbf{66.8}
  & \textbf{69.5} \\[0.5ex]

\adjustbox{valign=m}{\includegraphics[height=1.5em]{figures/icons/meta.png}}\quad Llama 4 Maverick
  & 53.1
  & 61.1
  & 64.2
  & \underline{69.3} \\[0.5ex]

\adjustbox{valign=m}{\includegraphics[height=1.5em]{figures/icons/deepseek.png}}\quad DeepSeek Chat v3
  & \underline{53.5}
  & 60.7
  & 63.7
  & 66.2 \\[0.5ex]

\adjustbox{valign=m}{\includegraphics[height=1.5em]{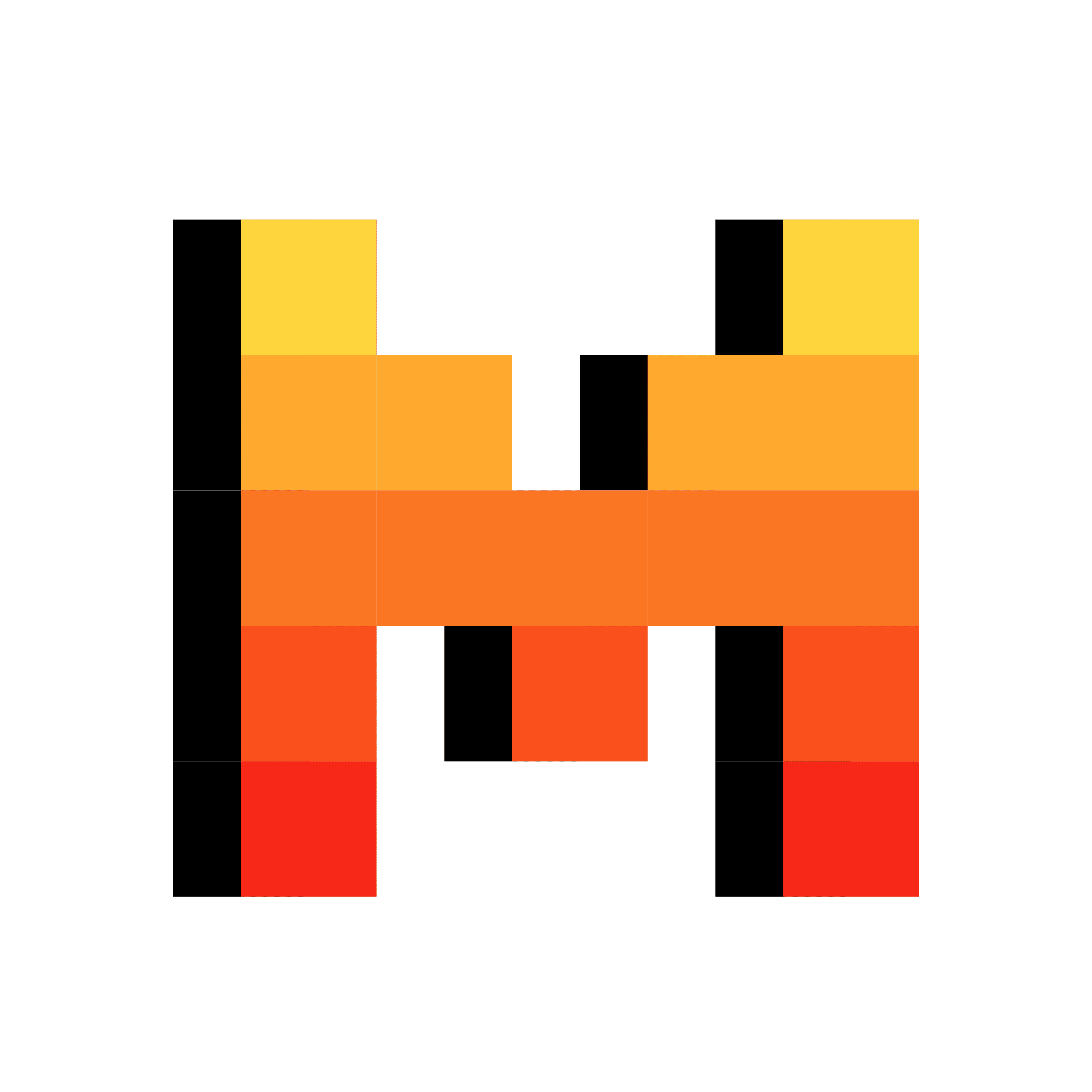}}\quad Mistral Small 3.1
  & 52.0
  & 59.7
  & 61.7
  & 63.9 \\[0.5ex]
  
\adjustbox{valign=m}{\includegraphics[height=1.5em]{figures/icons/deepmind.png}}\quad Gemini 2.0 Flash-001
  & 51.4
  & 59.2
  & 62.0
  & 66.2 \\[0.5ex]

\adjustbox{valign=m}{\includegraphics[height=1.5em]{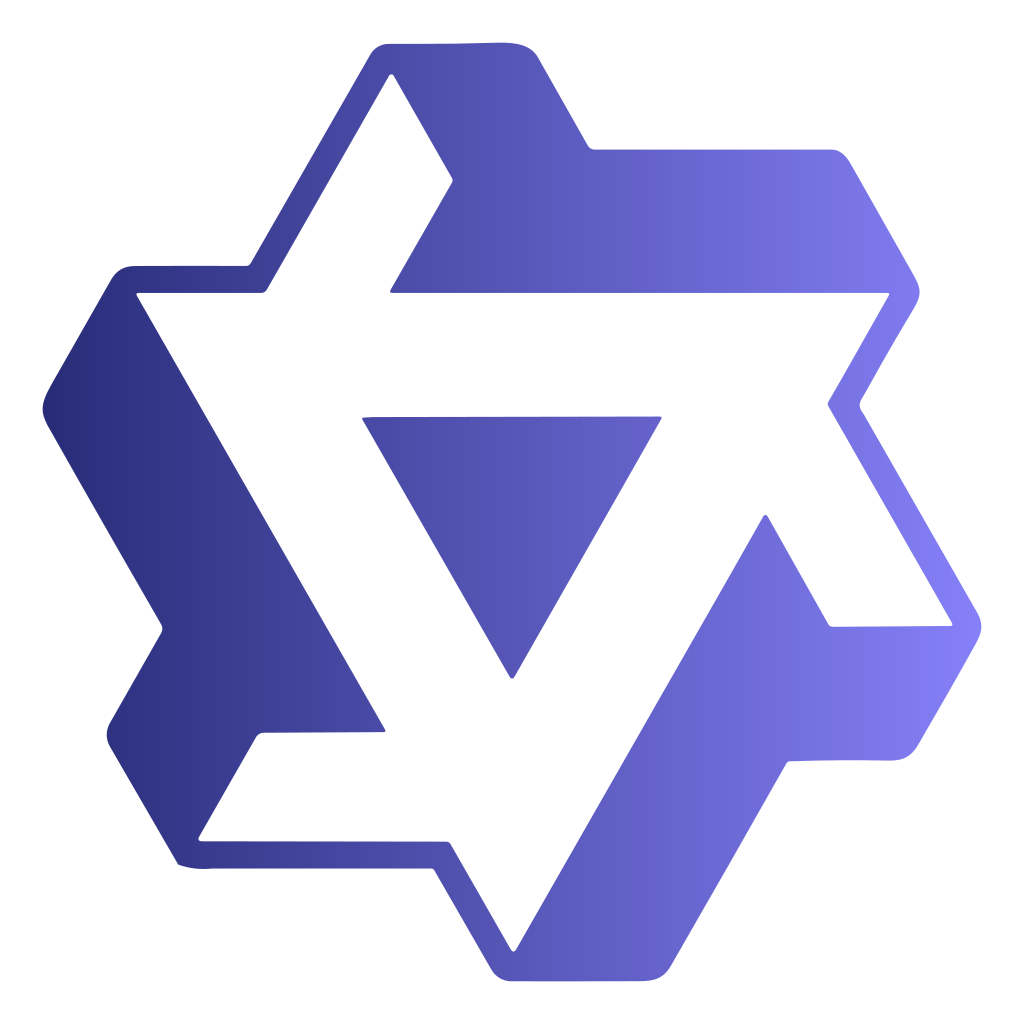}}\quad Qwen 2.5 VL
  & 50.7
  & 55.5
  & 58.0
  & 59.3 \\
\bottomrule
\end{tabular}
}
\endgroup
\end{table*}

\begin{figure}[htbp]
    \centering
    \includegraphics[width=1\textwidth]{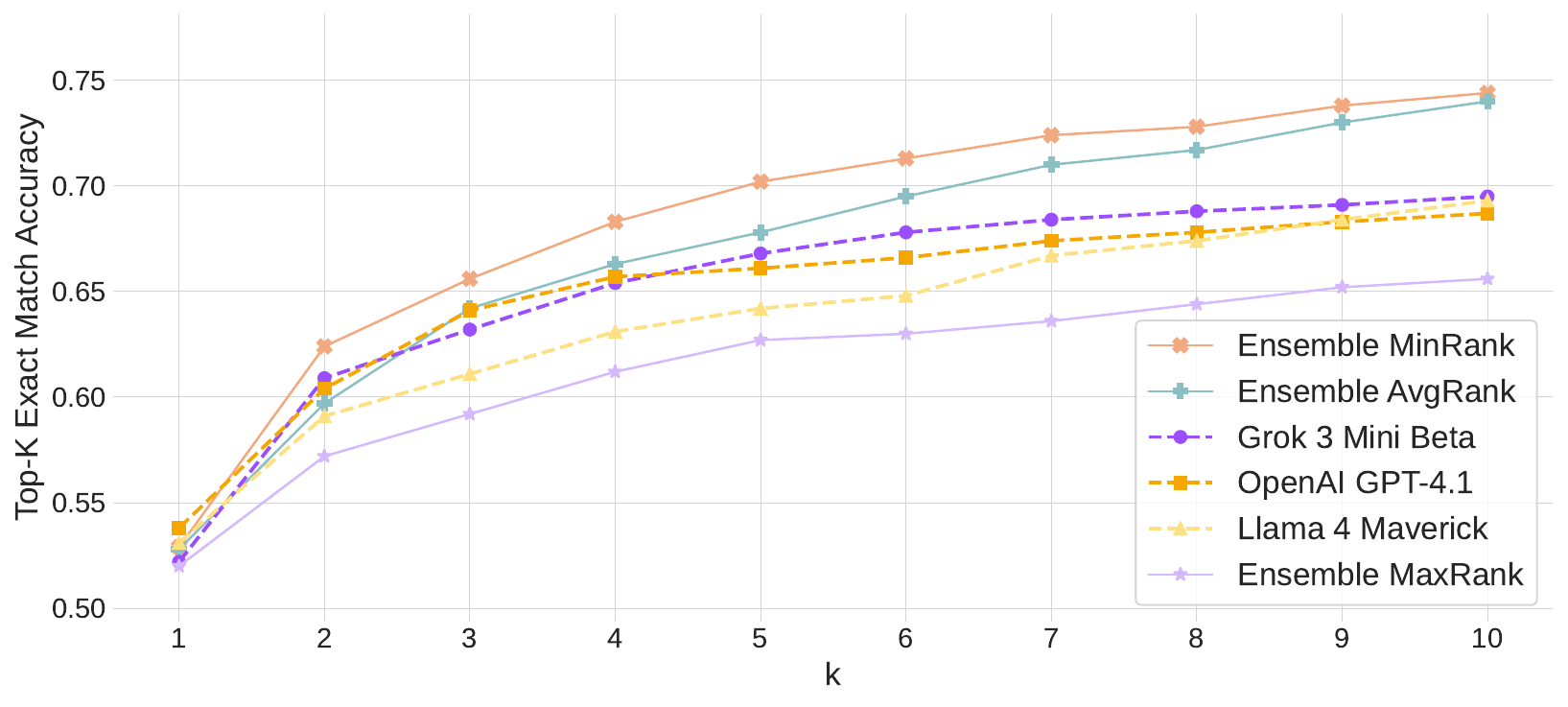} 
    \caption{\textbf{Ensemble Comparison.} Top-$k$ exact-match accuracy for three individual language models: Grok~3~Mini, GPT-4.1, and Llama~4~Maverick; and their joint ensemble with predictions combined using minimum-rank, average-rank, and maximum-rank voting. The minimum-rank ensemble achieves the best recall beyond Top–1.}

    \label{fig:ensemble_graph}
\end{figure}




\subsubsection{LM ensemble}
Finally, we evaluate the suitability of an ensemble approach. Using performance on the validation set, we construct an ensemble of LMs comprising Gemini 2.0 Flash, Llama 4 Maverick, and DeepSeek Chatv3, and compare three aggregation strategies:
\begin{itemize}
\item \textbf{Min-rank:} assign each item the best (lowest) rank it received across all models, promoting any item that at least one model ranks highly.

\item \textbf{Average-rank:} compute the average rank across models, balancing contributions from all models and reducing the impact of any single outlier.

\item \textbf{Max-rank:} assign each item the worst (highest) rank it received, ensuring only items consistently favored by every model appear at the top.
\end{itemize}

We observe that the Min-rank and Average-rank schemes substantially improve performance at Top-3, Top-5, and Top-10 (Fig. \ref{fig:ensemble_graph}), at the expense of a small drop in Top-1 accuracy compared to the best individual model. The high recall of our ensemble arises from its diversity. In the information-retrieval literature, rank-fusion methods consistently improve recall by leveraging the complementary strengths of diverse rankers across topics and queries, demonstrating that greater diversity correlates with enhanced recall in ranking tasks \citep{wu2020promoting, cormack2009reciprocal}. Consequently, min-rank and average-rank aggregation schemes are particularly effective for our LM ensembles: they ensure that any candidate ranked highly by a single model enters the final Top-$k$, thus maximizing recall.


\subsubsection{Synthesis condition regression}

For synthesis condition regression, we evaluate LM performance in predicting parameters for a standard solid-state synthesis protocol involving two heating steps. First, precursor powders are mixed and homogenized to ensure uniform distribution. Next, in the calcination stage, the precursor blend is calcined, activating thermal decomposition and diffusion that drive formation of the target phase. Finally, during the sintering stage, elevated temperature promotes atomic grain–boundary and volume diffusion, which drives neck formation and growth, thereby consolidating and densifying the powder particles into a cohesive bulk body \cite{kingery1976introduction,reed1995principles}.
In order to replicate experimental workflows, we prompt the language models to predict calcination and sintering temperatures. The generated conditions are then compared against the curated 1000 entry subset of the data published by Kononova \cite{kononova2019text}. We omit the associated dwell times here, as they are found to be strongly dependent on human preference and thus difficult to predict via regression \cite{karpovich2023interpretable, huo2022machine}. As a result, prior point‐regression models have tended to capture researcher noise rather than underlying thermodynamics, resulting in low $R^2$ and poor predictive performance \cite{karpovich2023interpretable, huo2022machine}.

Moreover, for regression tasks, it must be taken into account that language models are fundamentally next-token predictors trained on a classification objective. They lack native capabilities for precise numerical reasoning. As a result, they are ill-suited for regression tasks \cite{zausinger2024regress}. However, different works have focused on applying LMs to difficult numerical tasks with considerable success \cite{cobbe2021training}.
Furthermore, in practice synthesis temperatures are almost always reported as integer‐valued whole numbers (for example, 800 °C), which further eases the applicability of language models.

\begin{table*}[t]
  \centering
  \caption{\textbf{Synthesis condition prediction performance.} Regression performance for calcination and sintering temperature prediction. \textbf{Bold} indicates the best and \underline{underlined} the second-best value in each metric. Ensemble 1 comprises Gemini 2.0 Flash, Llama 4 Maverick, and DeepSeek Chat v3, Ensemble 2 features OpenAI GPT-4.1, Gemini 2.0 Flash, and DeepSeek Chat v3.}
  \label{tab:temp_performance}
  \begingroup
    \footnotesize
    \resizebox{\linewidth}{!}{%
      \begin{tabular}{l c c c @{\hskip 1.5em} c c c}
        \toprule
        \textbf{Model} 
          & \multicolumn{3}{c}{\textbf{Sintering temperature}} 
          & \multicolumn{3}{c}{\textbf{Calcination temperature}} \\
        \cmidrule(lr){2-4} \cmidrule(lr){5-7}
          & \textbf{MAE}~($\downarrow$) 
          & $\mathbf{R}^2$~($\uparrow$)
          & \textbf{RMSE}~($\downarrow$) 
          & \textbf{MAE}~($\downarrow$) 
          & $\mathbf{R}^2$~($\uparrow$)
          & \textbf{RMSE}~($\downarrow$) \\
        \midrule

        \raisebox{-0.3\height}{\includegraphics[height=1.5em]{figures/icons/meta.png}}\,%
        \raisebox{-0.3\height}{\includegraphics[height=1.5em]{figures/icons/deepmind.png}}\,%
        \raisebox{-0.3\height}{\includegraphics[height=1.5em]{figures/icons/deepseek.png}}%
        \quad Ensemble Avg
          & \textbf{96.31} 
          & \textbf{0.667} 
          & \textbf{134.48}
          & 125.72 
          & 0.410 
          & 168.86     \\[0.5ex]

        \raisebox{-0.3\height}{\includegraphics[height=1.5em]{figures/icons/openai.png}}\,%
        \raisebox{-0.3\height}{\includegraphics[height=1.5em]{figures/icons/deepmind.png}}\,%
        \raisebox{-0.3\height}{\includegraphics[height=1.5em]{figures/icons/deepseek.png}}%
        \quad Ensemble Avg
          & 96.89 
          & 0.6627 
          & 135.42  
          & \textbf{123.00} 
          & \textbf{0.424} 
          & \textbf{166.93} \\[0.5ex]
        \midrule

        \adjustbox{valign=m}{\includegraphics[height=1.5em]{figures/icons/deepmind.png}}%
        \quad Gemini 2.0 Flash-001
          & \textbf{100.66} 
          & \textbf{0.628} 
          & \textbf{142.22}
          & \underline{127.04} 
          & \underline{0.356} 
          & \underline{176.53} \\[0.5ex]

        \adjustbox{valign=m}{\includegraphics[height=1.5em]{figures/icons/meta.png}}%
        \quad Llama 4 Maverick
          & \underline{102.76} 
          & \underline{0.612} 
          & \underline{145.23}
          & 135.85            
          & 0.323             
          & 180.90          \\[0.5ex]

        \adjustbox{valign=m}{\includegraphics[height=1.5em]{figures/icons/openai.png}}%
        \quad OpenAI GPT-4.1
          & 105.21 
          & 0.586 
          & 150.01
          & \textbf{125.92} 
          & \textbf{0.371} 
          & \textbf{174.45} \\[0.5ex]

        \adjustbox{valign=m}{\includegraphics[height=1.5em]{figures/icons/deepseek.png}}%
        \quad DeepSeek Chat v3
          & 106.40 
          & 0.610 
          & 145.73
          & 132.48 
          & 0.309 
          & 182.78                       \\[0.5ex]

        \adjustbox{valign=m}{\includegraphics[height=1.5em]{figures/icons/mistral.png}}%
        \quad Mistral Small 3.1
          & 113.93 
          & 0.550 
          & 156.36
          & 137.05 
          & 0.291 
          & 185.20          \\[0.5ex]

        \adjustbox{valign=m}{\includegraphics[height=1.5em]{figures/icons/qwen.png}}%
        \quad Qwen 2.5 VL
          & 131.93 
          & 0.443 
          & 174.06
          & 142.68 
          & 0.232 
          & 192.72          \\[0.5ex]

        \adjustbox{valign=m}{\includegraphics[height=1.5em]{figures/icons/xai.png}}%
        \quad Grok-3 Mini Beta
          & 131.00 
          & 0.433 
          & 175.56
          & 152.09 
          & 0.123 
          & 205.97          \\

        \bottomrule
      \end{tabular}%
    }
  \endgroup
\end{table*}

Table \ref{tab:temp_performance} presents the results of our experiments, comparing the performance of LMs on the synthesis condition regression task. 
For calcination temperature regression, OpenAI~GPT-4.1 is the best performing model, followed by Gemini~2.0 Flash and DeepSeek Chat~v3. In sintering temperature regression, Gemini~2.0 Flash achieves the highest accuracy, with Llama~4 Maverick and OpenAI~GPT-4.1 ranking next. Grok~3 Mini Beta, previously second in precursor prediction, ranks lowest in both regression tasks. In general we see a correlation between model performance on both tasks.


Single‐model approaches achieve MAEs of 101~$^\circ$C in predicting sintering temperatures, compared with an MAE of 127~$^\circ$C for calcination temperature regression.
This is considerable respective the fact, that sintering temperatures are overall of higher magnitudes. 
We assess the source for the elevated error in calcination temperature predictions by benchmarking both the sintering and calcination tasks against a mean‐predictor baseline (Table \ref{tab:mean_predictor_baseline}). We find that the normalized standard deviation of calcination temperatures is roughly 16 \% higher than that of sintering temperatures (Fig.~\ref{histograms}), indicating a broader spread around the mean, thus accounting for the increased MAE in calcination predictions. Moreover, calcination temperature may have a stronger dependence on factors not reported in the dataset than sintering temperatures, such as variations in precursor particle size, a known factor in calcination conditions. For example, Pavlović \emph{et al.} report that extending ball-milling duration for BaTiO$_3$ by one hour reduces the required calcination temperature by over $100\,^\circ\mathrm{C}$ \cite{pavlovic2008synthesis}.

Similar to the precursor recommendation task, we evaluate the performance of an ensemble of LMs by taking the average of predictions of three LMs. For the sintering temperature regression, we use LLama 4 Maverick, Gemini 2.0 Flash and DeepSeek Chat v3. Thereby we see a distinct performance improvement of 4\% in $R^2$ over the best single-LM. For calcination temperature regression we add in a second ensemble by exchanging Llama with GPT-4.1. Notably this setup is capable of increasing calcination temperature $R^2$ values by 5\% to a reasonable 42.4\%.  Again, in close agreement with the precursor recommendation task, ensemble model configurations reduces the inference cost by around 70\%, while boosting performance.

\begin{figure}[tbp]
  \centering
  \includegraphics[width=\linewidth]{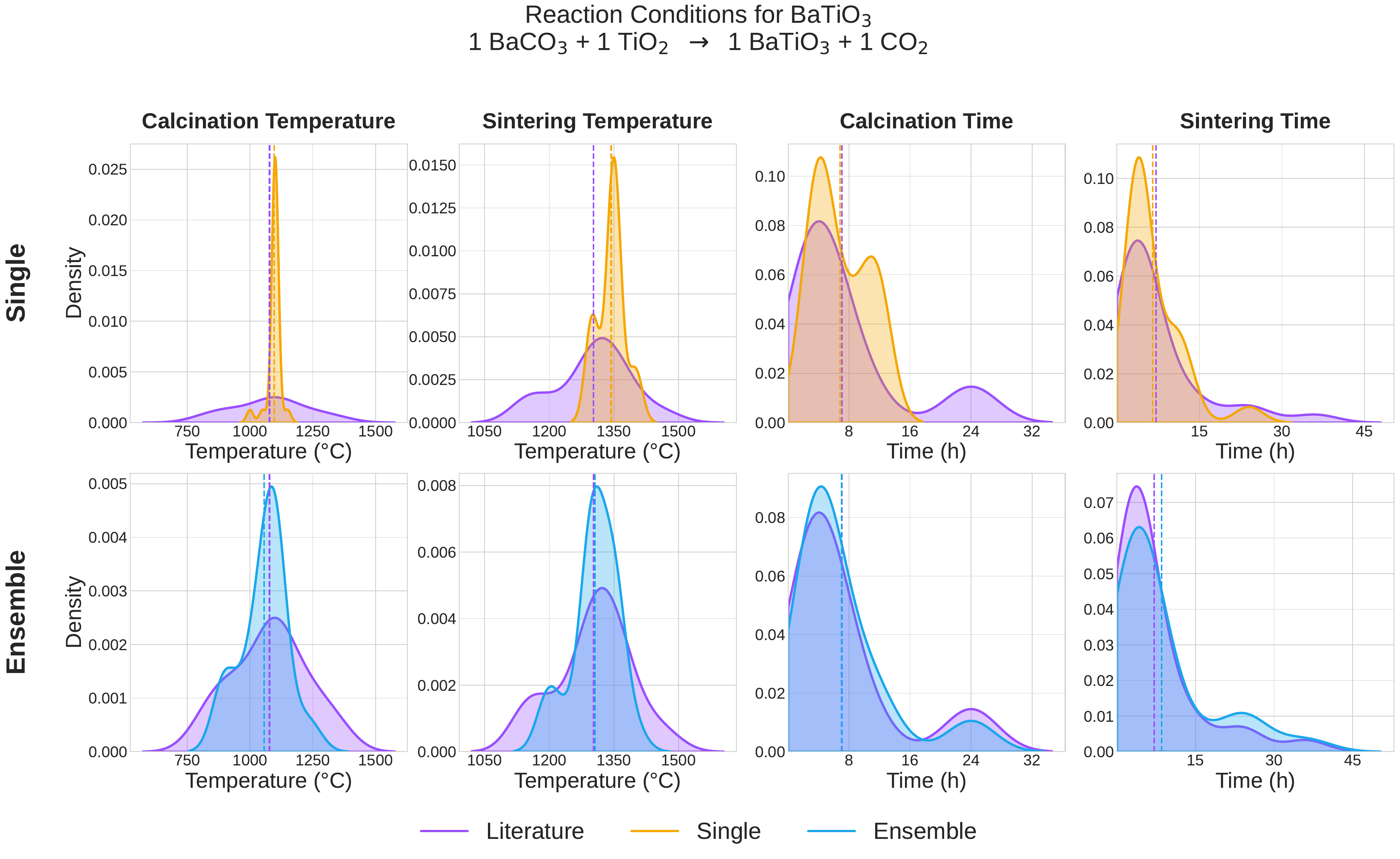}
  \caption{\textbf{Synthesis condition distributions of literature-reported and LM generated solid-state synthesis recipes for BaTiO\textsubscript{3}.} Literature distribution KDEs are shaded purple. Dotted lines refer to the mean value. 'Single' refers to LM distributions acquired by drawing 24 samples from Gemini-2.0-Flash (orange). 'Ensemble' refers to LM distributions acquired by drawing 8 predictions each are sampled from Llama Maverick, DeepSeek Chat v3, and Gemini 2.0 Flash (blue). The individual LMs yield narrower distributions that fail to capture the underlying literature distribution, whereas the ensemble more accurately reproduces the literature’s secondary modes and long-tail behavior.}
  \label{fig:reaction-conditions-batio3}
\end{figure}

We rationalize the underlying reason why a LM ensemble outperforms individual LMs. In materials synthesis, the mapping from processing recipes to target materials is inherently one-to-many\cite{pan2024chemically}. A single composition, such as BaTiO$_3$, can be produced through multiple annealing protocols that vary in calcination and sintering conditions, most notably in temperature and dwell time. We LM-generate a distribution of synthesis conditions and compare to the literature-reported synthesis of 24 pristine BaTiO$_3$ samples. As shown in Fig. \ref{fig:reaction-conditions-batio3}, prompting individual LMs yields narrow distributions with a single dominating mode peaked sharply around the means (orange, upper row). A LM-ensemble achieves substantially improved overlap with the ground-truth (purple). For example, in the case of calcination temperature, it correctly predicts a secondary mode below the mean while accurately reproducing the primary mode (blue, lower row). Similarly, for sintering temperature, the ensemble distribution captures the mean and additional features of the target distribution, such as a regime near 1200\,\textdegree C. Most notably, for synthesis time, individual LMs regress to the mean with a narrow spread while the LM-ensemble more accurately captures the long tail distribution at longer processing durations. As such, LM-ensembles better capture one-to-many target- synthesis relationships, which offers a key insight into why they outperform individual LMs.

\subsection{Performance vs. cost tradeoff}

\begin{figure}[t]
    \centering
    \includegraphics[width=1\textwidth]{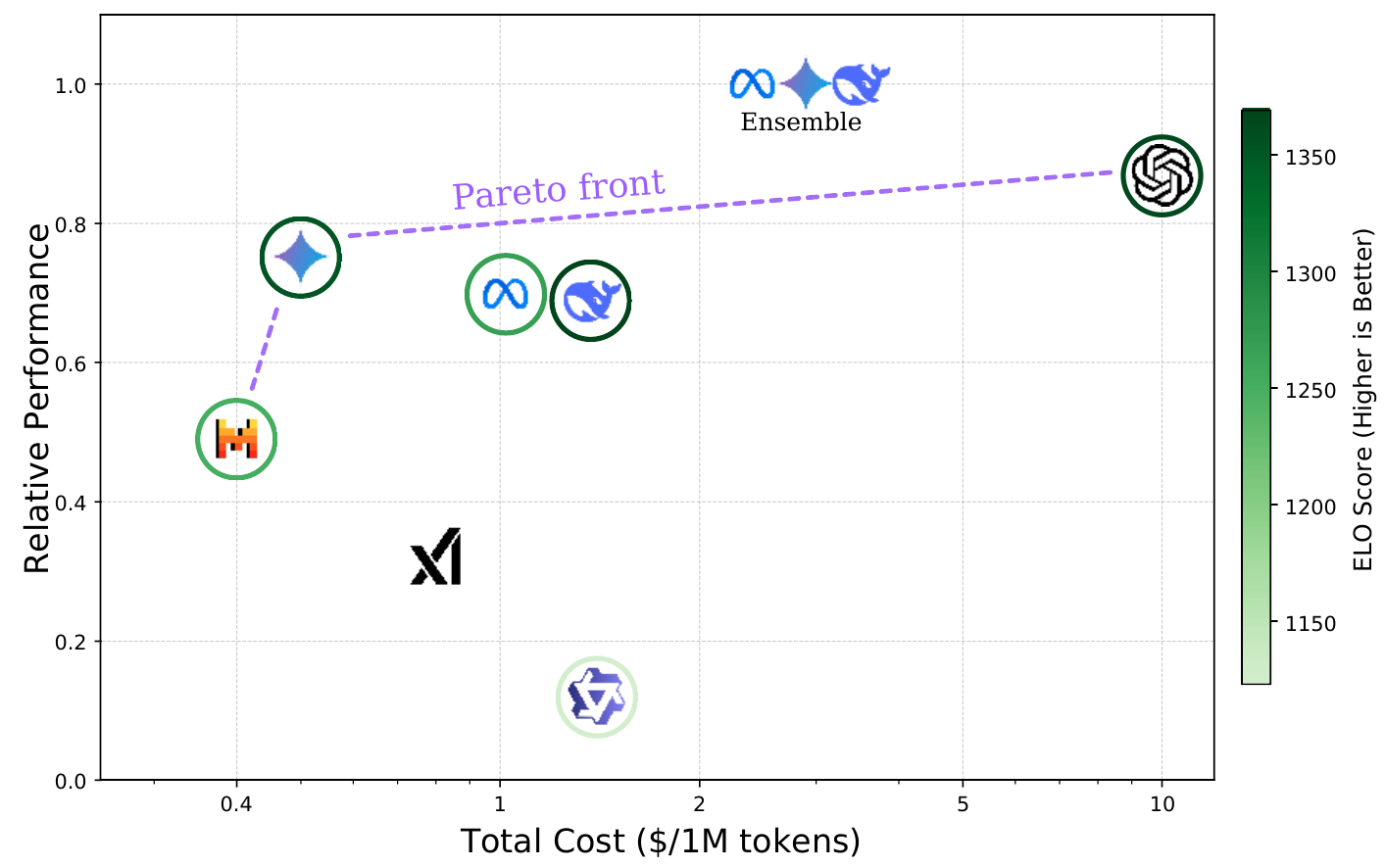}
    \caption{\textbf{Comparison of model performance vs. cost} We compute each model’s relative performance on precursor prediction, calcination and sintering temperature estimation tasks, and plot the average performance relative to cost. GPT-4.1 delivers highest individual performance and comes at the highest cost. An ensemble of Llama~4 Maverick, DeepSeek Chat v3 and Gemini 2.0 Flash surpasses any single model in performance while reducing cost by 70\% relative to GPT-4.1. The Elo rating score is represented by the color of each circle and serves as a quantitative indicator of model performance \cite{chiang2024chatbot}.}
    \label{fig:model_cost_vs_performance}
\end{figure}

To compare the overall LM performances, we normalize each model’s performance to that of the best-performing model and compute the mean normalized score. We estimate costs using input and output token rates (Fig. \ref{fig:model_cost_vs_performance}). GPT-4.1 and Gemini achieve the highest average performance. However, Gemini’s substantially lower price point makes it especially attractive for materials-informatics applications. Notably, the ensemble of lower-priced models: Llama~4 Maverick, DeepSeek Chat v3 and Gemini 2.0 Flash outperforms any single model while reducing cost by 70\% relative to the top performing GPT-4.1. Moreover, as shown by our analysis, ensembles yield output distributions more closely aligned with scientific literature, underscoring the joint cost and performance benefits.

\newpage

\section{Augmenting solid-state synthesis prediction with synthetic data}
Our previous investigation demonstrated that LMs achieve strong performance in recalling already published synthesis parameters for solid-state synthesis. 
This result opens up the potential of leveraging LMs to generate synthetic datasets over synthesis parameter distributions to augment size-limited experimental datasets.

\subsection{Generating a diverse dataset}

\begin{figure}[t]
    \centering
    \includegraphics[width=1\textwidth]{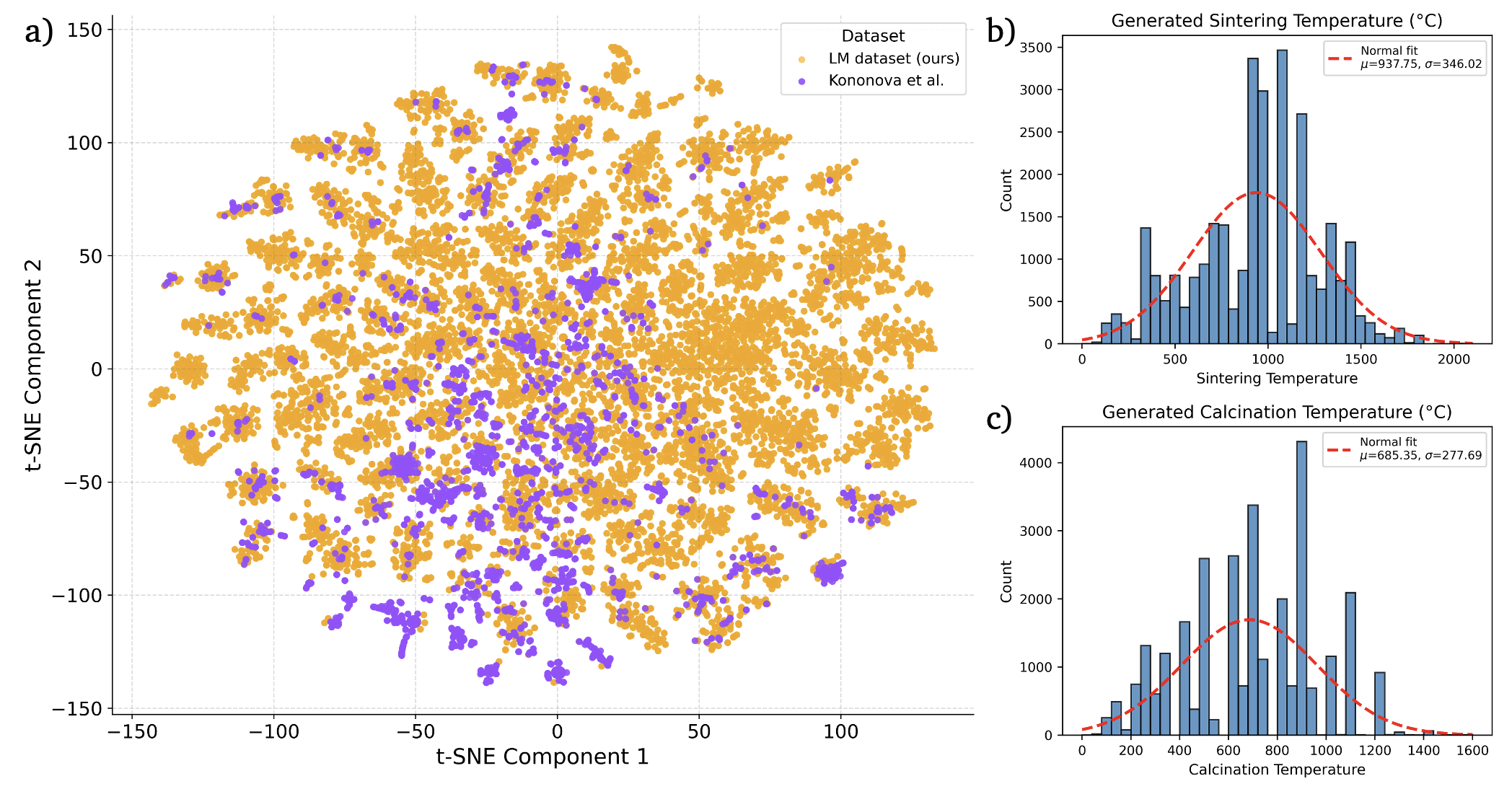}
    \caption{\textbf{t‐SNE projection of inorganic precursor compositions in embedding space.} Compositions from the Kononova dataset \cite{kononova2019text} are shown in blue, and LM‐generated compositions in orange. Compositions are represented by standardized elemental‐fraction vectors and projected via t‐SNE. The expanded spread of orange points indicates that our generated dataset spans a larger chemical composition space than the baseline dataset. Distributions on the right are reported before filtering by temperature threshold.}
    \label{fig:gen_dataset_fig}
\end{figure}

To assemble a diverse dataset, we queried the Materials Project \cite{jain2013commentary}, yielding 48,927 lab-synthesized compounds. We then applied maximum-entropy sampling to select 10,000 target compositions, thereby ensuring maximal compositional diversity. Using GPT–4.1, we generate precursor routes and prompt the model to flag and remove materials not synthesizable via solid–state methods. In line with previous findings, we preserve the top three predictions per material, given the model’s robust top–3 accuracy of 64.1\% (Table \ref{tab:model_performance_llm_precursor}). We then predict synthesis parameters, producing 29,473 entries, excluding generated routes found to be incomplete. Incorporating minimum temperature thresholds of 300,\textdegree{}C for calcination and 500,\textdegree{}C for sintering yields 28,548 plausible solid-state recipes. Figure \ref{fig:gen_dataset_fig}a) shows the enhanced compositional diversity of the generated dataset, when compared with the literature-mined dataset by Kononova \cite{kononova2019text}.

\subsection{Synthetic data augmentation improves model performance} 
\begin{figure}[htbp]
    \centering
    \includegraphics[width=0.9\textwidth]{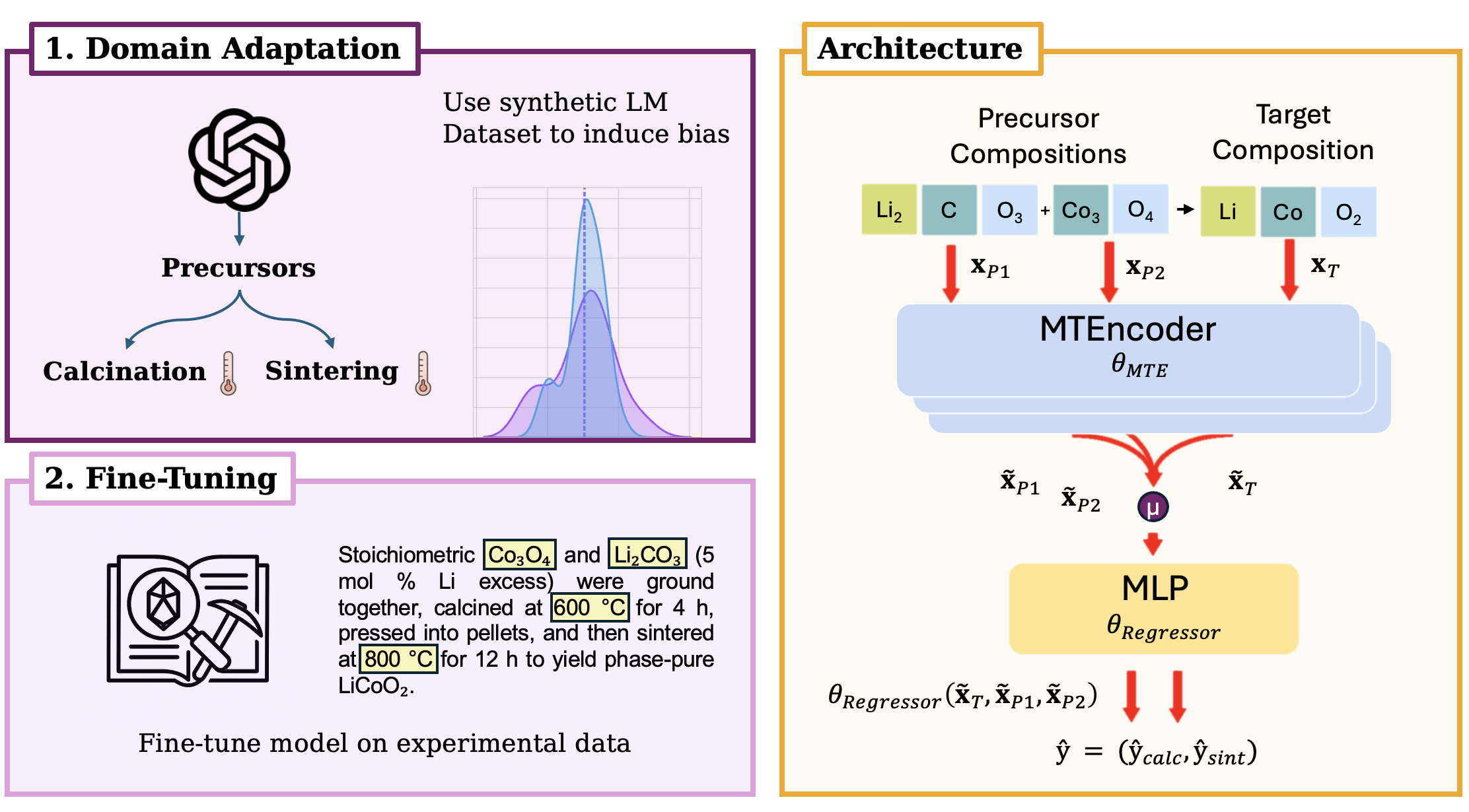} 
    \caption{%
    \textbf{Overview of our synthesis‐condition modeling.} \textit{Left}: We first adapt the MTEncoder on a large LM‐generated dataset to bias it toward solid‐state reaction conditions, then fine-tune on experimental literature recipes. \textit{Right}: Each precursor and the target composition are encoded by the shared MTEncoder ($\theta_{\mathrm{MTE}}$) into embeddings $\tilde x_{p,i},\tilde x_T$, pooled and concatenated, and passed through an MLP head ($\theta_{\mathrm{Regressor}}$) to predict calcination and sintering temperatures $\hat y=(\hat y_{\mathrm{calc}},\hat y_{\mathrm{sint}})$.%
    }
    \label{fig:flow_syntmte}

\end{figure}

Following generation of our LM-derived dataset, we evaluated its impact on synthesis-condition regression models. This approach mirrors NLP strategies that augment scarce domain-specific data with LM-generated corpora (e.g., Xu et al.\ \cite{xu2023s2ynre}) or employ teacher–student pseudo-labeling frameworks \cite{valizadehaslani2024two,pieper2024enhancing,tran2025domain}.
Incorporating this workflow, we propose \texttt{SyntMTE}, a composition-based architecture derived from MTEncoder, a transformer model for representing inorganic materials, pretrained on the large Alexandria DFT database \cite{prein2023mtencoder, schmidt2024improving}. 
Our approach extends previous work \cite{karpovich2023interpretable} by encoding both the reaction products and all associated precursors into embedding vectors. Subsequently, they are aggregated via mean pooling (Fig.~\ref{fig:flow_syntmte}, right), and process parameters are predicted via multi‐task regression. In ablation, mean and sum pooling outperform learned LSTM and sequential aggregation schemes (Table~\ref{tab:ablation_pooling}). We benchmark on the Kononova dataset \cite{kononova2019text} against three baselines, a compositional feedforward network, a CrabNet-based transformer \cite{wang2021compositionally}, and an XGBoost regressor with mean-pooled reaction features. The models are compared on three training regimes: a two‐stage fine‐tuning on synthetic and subsequently literature data, direct fine‐tuning on literature data only, and training exclusively on synthetic data.

\begin{table}[htb]
  \centering
    \caption{%
      \textbf{Model comparison.} Comparison of embedding methods on different data regimes for sintering and calcination temperatures. We report mean values across five runs with standard deviation in parentheses. \checkmark\ indicates training on the data source.
    }
  \label{tab:synthetic_data_augmentation}
  \renewcommand{\arraystretch}{0.9}
  \scriptsize
  \setlength{\tabcolsep}{5pt}
  \begin{adjustbox}{width=\textwidth,center}
    \begin{tabular}{@{} 
        l 
        cc 
        *{3}{c} 
        *{3}{c}
        c 
      @{} }
      \toprule
      & \makecell{\textbf{Synth.}\\\textbf{Data}} 
      & \makecell{\textbf{Literature}\\\textbf{Data}} 
      & \multicolumn{3}{c}{\textbf{Sintering temperature}} 
      & \multicolumn{3}{c}{\textbf{Calcination temperature}}
      & \textbf{Rel. MAE Imp.} \\
      \cmidrule(lr){4-6} \cmidrule(lr){7-9}
      \textbf{Model} 
        &  &  
        & \textbf{MAE $\downarrow$} & \textbf{RMSE $\downarrow$} & \textbf{R$^2$ $\uparrow$}
        & \textbf{MAE $\downarrow$} & \textbf{RMSE $\downarrow$} & \textbf{R$^2$ $\uparrow$}
        & \textbf{$\uparrow$} \\
      \midrule
      \texttt{SyntMTE}
        & \checkmark & \checkmark
        & \makecell{\textbf{72.94}\\(0.85)}
        & \makecell{\textbf{131.37}\\(1.14)}
        & \makecell{0.616\\(0.007)}
        & \makecell{\textbf{98.39}\\(0.36)}
        & \makecell{\underline{155.80}\\(0.39)}
        & \makecell{0.423\\(0.003)}
        & \impGrad{3.68} \\
      \texttt{SyntMTE}
        &            & \checkmark
        & \makecell{\underline{74.82}\\(1.27)}
        & \makecell{\underline{133.48}\\(1.88)}
        & \makecell{0.603\\(0.011)}
        & \makecell{103.05\\(1.64)}
        & \makecell{163.63\\(3.87)}
        & \makecell{0.363\\(0.030)}
        & \impGrad{0.00} \\
      \texttt{SyntMTE}
        & \checkmark &            
        & \makecell{88.54\\(1.27)}
        & \makecell{150.17\\(2.13)}
        & \makecell{\underline{0.619}\\(0.011)}
        & \makecell{112.98\\(2.83)}
        & \makecell{168.65\\(3.61)}
        & \makecell{\textbf{0.502}\\(0.021)}
        & \impGrad{-13.30} \\
      \midrule
    CrabNet\cite{wang2021compositionally}
      & \checkmark & \checkmark
      & \makecell{75.11\\(0.90)}
      & \makecell{134.28\\(1.36)}
      & \makecell{0.599\\(0.008)}
      & \makecell{\underline{99.11}\\(1.00)}
      & \makecell{\textbf{153.78}\\(0.90)}
      & \makecell{0.438\\(0.007)}
      & \impGrad{8.71} \\
    CrabNet\cite{wang2021compositionally}
      &            & \checkmark
      & \makecell{84.75\\(2.27)}
      & \makecell{149.06\\(4.52)}
      & \makecell{0.505\\(0.030)}
      & \makecell{106.89\\(1.21)}
      & \makecell{164.11\\(1.16)}
      & \makecell{0.359\\(0.009)}
      & \impGrad{0.00} \\
    CrabNet\cite{wang2021compositionally}
      & \checkmark &            
      & \makecell{90.06\\(2.05)}
      & \makecell{143.48\\(3.12)}
      & \makecell{\textbf{0.627}\\(0.016)}
      & \makecell{111.80\\(2.22)}
      & \makecell{169.51\\(2.60)}
      & \makecell{\underline{0.475}\\(0.016)}
      & \impGrad{-5.33} \\
      \midrule
      Composition + NN
        & \checkmark & \checkmark
        & \makecell{77.32\\(1.44)}
        & \makecell{136.38\\(3.09)}
        & \makecell{0.586\\(0.019)}
        & \makecell{101.91\\(1.58)}
        & \makecell{159.25\\(2.59)}
        & \makecell{0.397\\(0.020)}
        & \impGrad{5.74} \\
      Composition + NN
        &            & \checkmark
        & \makecell{83.68\\(2.27)}
        & \makecell{145.69\\(2.89)}
        & \makecell{0.527\\(0.019)}
        & \makecell{106.46\\(1.29)}
        & \makecell{165.98\\(1.24)}
        & \makecell{0.345\\(0.010)}
        & \impGrad{0.00} \\
      Composition + NN
        & \checkmark &            
        & \makecell{94.62\\(3.63)}
        & \makecell{158.10\\(4.41)}
        & \makecell{0.577\\(0.024)}
        & \makecell{123.48\\(1.50)}
        & \makecell{182.30\\(1.57)}
        & \makecell{0.418\\(0.010)}
        & \impGrad{-14.71} \\
      \midrule
    Composition + XGBoost
      &            & \checkmark
      & \makecell{126.19\\(1.38)}
      & \makecell{175.12\\(1.25)}
      & \makecell{0.584\\(0.006)}
      & \makecell{135.99\\(1.62)}
      & \makecell{182.26\\(0.91)}
      & \makecell{0.422\\(0.006)}
      & \impGrad{0.00} \\
    Composition + XGBoost
  & \checkmark & 
  & \makecell{149.49\\(1.21)}
  & \makecell{196.99\\(1.46)}
  & \makecell{0.473\\(0.008)}
  & \makecell{162.09\\(1.33)}
  & \makecell{207.42\\(1.77)}
  & \makecell{0.251\\(0.013)}
  & \impGrad{-18.83} \\

          \bottomrule
    \end{tabular}
  \end{adjustbox}
\end{table}

As shown in Table~\ref{tab:synthetic_data_augmentation}, \texttt{SyntMTE} achieves the lowest MAEs across both tasks evaluated, followed by CrabNet and the compositional neural network. When trained exclusively on literature-mined experimental data, \texttt{SyntMTE} attains a MAE that is approximately $9\,^\circ\mathrm{C}$ lower for sintering and reduces the MAE by over $3\,^\circ\mathrm{C}$ for calcination compared to the baseline models. This indicates, that \texttt{SyntMTE}’s DFT pretraining objective provides a robust inductive bias, delivering competitive results even on small experimental datasets. Notably, when training on literature data only, the lightweight compositional neural network matches the performance of the parameter-intensive Transformer architecture in CrabNet. Although CrabNet is generally considered the more powerful model, its performance parity with the smaller architecture is attributable to the limited training set of only 3,400 distinct compositions, which constrains the benefits of greater model capacity and predisposes larger architectures to overfitting. Despite previous reports of strong performance \cite{karpovich2023interpretable}, XGBoost demonstrates the poorest scores among all models. We attribute this to the increased complexity of our task, in which the model receives an entire reaction as input instead of a target mixed with physical features \cite{karpovich2023interpretable}.

When trained exclusively on synthetic data, all models exhibit strong performance despite having no exposure to literature-mined data and being trained only on recipes that do not overlap with the literature-based test set. Notably, the $R^2$ value, which measures the proportion of variance in the target variable explained by the model remains high across all models. CrabNet achieves the highest $R^2$ values, with 62.7\% in sintering, while \texttt{SyntMTE} scores highest in calcination temperature $R^2$ 50.2\%, exceeding even $R^2$ values observed under the dual pretraining objective. We attribute this unexpectedly high performance to the model’s ability to capture general trends accurately via minimizing extreme outliers in its predictions. While augmenting the training set with experimental measurements reduces the sum of individual prediction errors, it simultaneously seems to introduce new occasional larger outliers, leaving the overall $R^2$ essentially unchanged for sintering and increased for calcination temperatures (Fig. \ref{fig:parity_r2}). Furthermore, the larger 29,457 entry dataset enables CrabNet to leverage its high parameter count and outperform the compositional feedforward network by a substantial margin.


Finally, we evaluate performance under both fine-tuning regimes. Across all models, augmenting the training set with synthetic data consistently enhances performance as evidenced by the relative MAE improvements in Table~\ref{tab:synthetic_data_augmentation}. This effect is most pronounced for CrabNet, which achieves a MAE improvement exceeding 8.7\% across both regression tasks. MTEncoder, by contrast, attains only a 3.7\% improvement, reflecting its stronger baseline performance from comprehensive DFT pretraining. The compositional feedforward network likewise realizes improvements exceeding 5.7\%, approaching the performance of the experimental data-trained \texttt{SyntMTE} model.
When comparing the scores to those of the best ensembles in the LM benchmark (see Table~\ref{tab:temp_performance}), the expert models continue to demonstrate superior performance across all regression tasks, exhibiting improvements of approximately 24\% in sintering and 20\% in calcination temperature MAE. Since LM benchmark scores may be inflated by data leakage, we ascribe their lower performance to the inherent challenges language models face in regression tasks.

Overall, comparison of models trained on literature‐mined versus synthetic data reveals a significant opportunity to leverage synthetic datasets in synthesis modeling. Training exclusively on large, LM‐generated synthetic datasets can achieve comparable performance to training on literature data, while eliminating the need for laborious manual data extraction. However, we also observe that the DFT-based pretraining objective of \texttt{SyntMTE} substantially improves performance when only limited literature data is available.

\section{Application to processing of  LLZO electrolyte materials}

\begin{figure}[h!]
  \centering
  \includegraphics[width=0.9\textwidth]{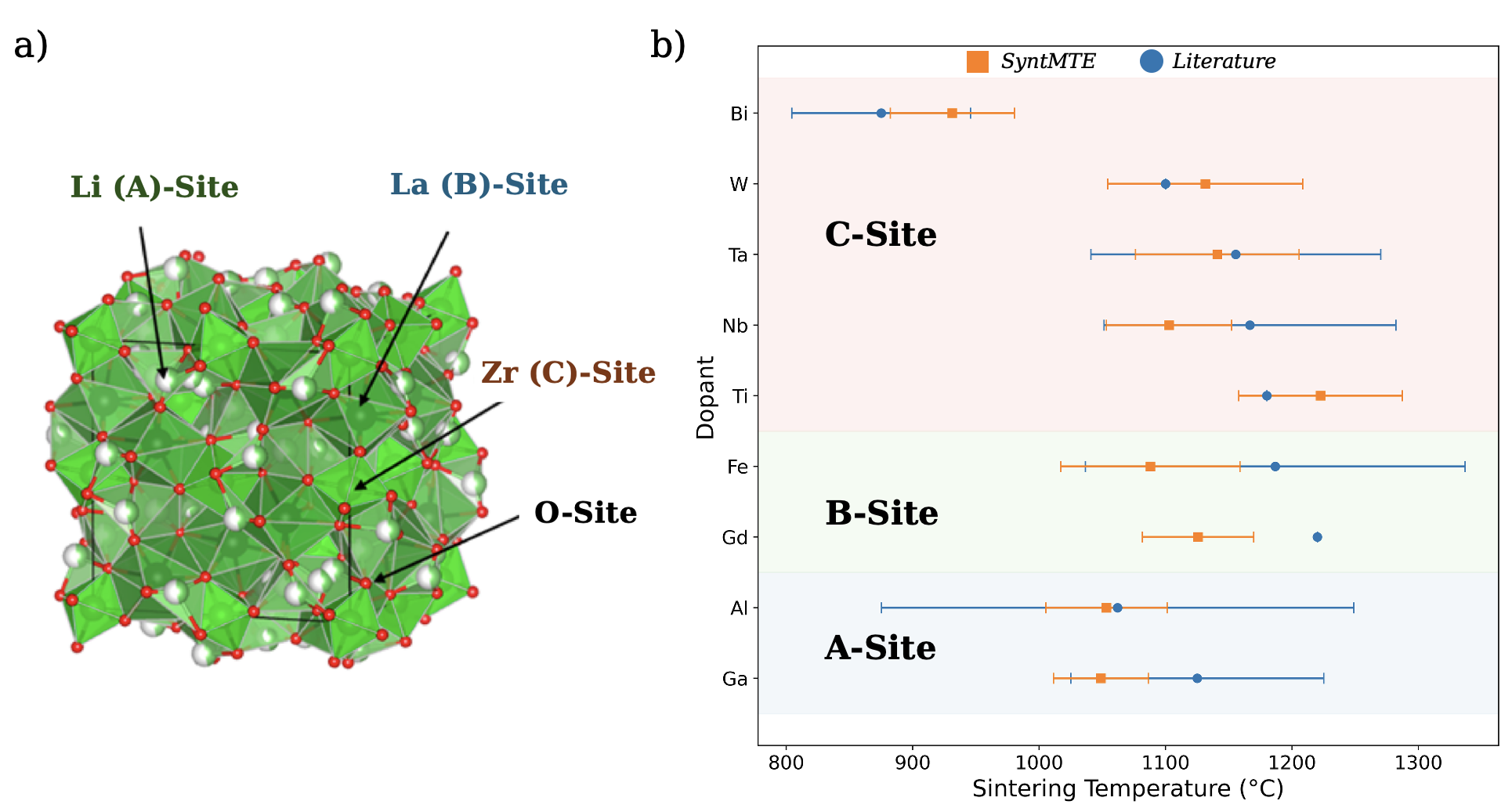}
  \caption{a) Illustrates the probable doping sites in the cubic LLZO unit cell, reproduced from \cite{mahbub2020text}. b) True (blue circles) vs.\ predicted (orange squares) sintering temperatures with mean and standard deviation across different case reports per cation of the garnet electrolyte. Dopants are grouped by their crystallographic substitution site.}
  \label{fig:case_study}
\end{figure}

Beyond assessing conventional performance-related material properties, virtual screening of compound-specific processing temperatures and durations offers a quantitative proxy for estimating manufacturing costs \cite{bauer2022charging}.
As a case study, we consider the processing of solid-state electrolytes, which function as the replacement for conventional liquid electrolytes in next generation hybrid and future solid-state Li-ion batteries \cite{balaish2021processing, Balaish2025Emerging}. Their key performance metrics are ionic conductivity and the electrochemical stability window \cite{kim2021solid}. Although oxide‐based electrolytes typically outperform alternatives in those metrics, they require densification through sintering at elevated temperatures, when processed in the form of free-standing electrolytes (tape or pellet)\cite{sand2024critical}. One of the most promising material candidates among solid state electrolytes is the garnet-type solid electrolyte Li$_7$La$_3$Zr$_2$O$_{12}$ (LLZO), which exhibits conductivities on the order of $10^{-3}\mathrm{ Scm^{-1}}$ at room temperature. However, widespread integration into next-generation battery architectures is still hindered by high processing costs, which originate from precursor selection and the sintering protocols required to fabricate electrolyte tapes \cite{Weinmann2025Sustainable}. Densification of cubic LLZO typically demands sintering at temperatures above $1,050^{\circ}\mathrm{C}$ for several hours, together with the incorporation of extrinsic phase-stabilising dopants\cite{mahbub2020text, Weinmann2025Sustainable, hood2022sinter, pfenninger2019low}. Consequently, one of the most pressing challenges is to reduce the sintering temperature, a common requirement for high-value functional ceramics\cite{mahbub2020text}. Studies have demonstrated that aliovalent doping at the Li (A), La (B) and Zr (C) sites can lower the sintering temperature while stabilizing the desired cubic phase (Figure \ref{fig:case_study}, a) \cite{Chu2025_GB_LLZO, mahbub2020text}.

This interesting case provides an opportunity to evaluate the resolution of our method for predicting solid-state sintering temperatures. We curate a dataset comprising 40 reported solid-state synthesis routes of doped LLZO variants in order to test if \texttt{SyntMTE} recovers compounds synthesizable at lower sintering temperatures. To test the extrapolation capabilities of our model, we withhold any LLZO-based compounds from the training sets and predict the sintering temperatures of various doped LLZO compositions.
Because literature‐reported distributions are difficult to reconstruct precisely, and several viable sintering temperatures exist per dopant family, we focus our analysis on the qualitative ordering of the compounds rather than on their exact values.


Among the C-site dopants investigated, tantalum is a well-studied yet comparatively moderate densification aid (Figure \ref{fig:case_study}, a).  
Super-valent substitution of $\mathrm{Ta^{5+}}$ for $\mathrm{Zr^{4+}}$ is charge-balanced by lithium vacancies ($V_{\mathrm{Li}}$), defect chemistry that accelerates lattice diffusion and stabilizes the cubic garnet framework \cite{li2024understanding, morozov2022thermodynamics, han2016electrochemical}.  
Experimentally, Ta-doped LLZO pellets are sintered for a few hours at $1100$–$1150^\circ\mathrm{C}$, about $100^\circ\mathrm{C}$ lower than the temperature needed for undoped LLZO and intermediate between the behaviors of W- and Bi-doped compositions \cite{kim2020role}. Our model accurately reproduces the literature synthesis temperature window, predicting a mean temperature closely matching reported values and a standard deviation reaching approximately from 1060\,\textdegree{}C to 1200\,\textdegree{}C.

Substitution of $\mathrm{Bi^{3+}}$ with  $\mathrm{Zr^{4+}}$-sites  induces significant lattice expansion through the larger ionic radius of  $\mathrm{Bi^{3+}}$ and yields higher lithium concentrations in the lattice, thereby enhancing ionic conductivity through larger migration channels, pronouncing the reduction of sintering temperature reported for a single‑site dopant \cite{schwanz2020bi,wagner2016synthesis}.  Our model predicts a narrow sintering window for Bi of $880-980^\circ\mathrm{C}$, overestimating the minimum slightly yet accurately capturing the sharp decline to approximately $900^\circ\mathrm{C}$ observed experimentally.  
For results at A- and B-site dopants, we observe Al spanning a broad sintering window owing to its small ionic radius and mixed-site occupancy, which promote gradual lattice relaxation and progressive vacancy clustering, thus smoothing the onset temperature \cite{kovsir2024supervalent, samson2019bird}. In contrast, Ga whose ionic radius is moderately larger than native lattice cations yields a narrower sintering profile. Our model’s prediction for Gd exhibits the largest deviation, being overly optimistic, whereas that for Fe aligns well with experimental values.

Overall, the results show that \texttt{SyntMTE}, despite receiving no prior training on LLZO, reproduces the key sintering temperature trends, underscoring the potential of synthesis planning models to guide future compound selection via virtual screening of processing temperatures.


\section{Conclusions}

Existing methods in machine learning-based materials synthesis planning remain limited by the available training data.
We demonstrate that current language models can overcome this shortcoming. We benchmark seven state-of-the-art models on two standard tasks: precursor recommendation and processing-parameter regression. Models such as GPT-4.1, Gemini-2.0, and Llama 4 Maverick achieve top-1 exact-match accuracies for precursor prediction above 50\%, rising to approximately 66\% for their top five suggestions. We find ensembles of language models enhancing performance further by accurately capturing the synthesis windows, meaning the multitude of processing conditions enabling the synthesis of the same target compound. In contrast, individual language models typically yield narrower, unimodal distributions, which reflect the synthesis window less accurately but yield comparably good results in regression tasks. Furthermore, ensembles can reduce inference costs by as much as 70\%.We then employ language models to distill materials-related knowledge into a synthetic solid-state synthesis dataset containing nearly 28,548 complete recipes. By including time and temperature information for every recipe, the resulting dataset is more than sixfold larger than existing text-mined databases. To quantify the utility of this synthetic augmentation, we develop \texttt{SyntMTE}, a transformer fine-tuned in two steps, first on the synthetic LM data and second on literature-based extracted data. \texttt{SyntMTE} outperforms existing baselines, including state-of-the-art CrabNet. Compared to training on experimental data only, our 2-step training enhances performance, by reducing MAEs of sintering and calcination temperature by $2~^\circ\mathrm{C}$ and $5~^\circ\mathrm{C}$, respectively. Overall, our experiments employing language-model-generated synthetic data demonstrate substantial potential for advancing research that leverages such datasets in inorganic synthesis.
In a case study, we apply our framework to doped variants of \ce{Li7La3Zr2O12} (LLZO), a solid-state electrolyte whose scalability is limited by an energy-intensive sintering process. Without LLZO-specific training, our model reproduces the broad sintering windows and captures detailed processing effects. This includes the substantial reduction in sintering temperature achieved by Bi substitution. These results demonstrate the model’s potential to identify low-temperature processing routes, for example, during the screening of novel compounds.

Collectively, our study demonstrates that language-model-based methods can generate high-quality, cost- and time-efficient auxiliary data on readily reported parameters and phenomena throughout the inorganic materials synthesis literature. This capability is critical as data remains scarce across the domain. Furthermore, by leveraging the generated data as pretraining, one can boost utilization of the limited lab results available. Ultimately such models can inform Bayesian optimization and guide autonomous experimentation, thereby accelerating the discovery and scalable production of advanced materials.

\newpage
\newpage

\section{Code Availability}
The source code underlying this project is available at the GitHub repository 
\href{https://github.com/Thorben010/llm_synthesis}{\url{https://github.com/Thorben010/llm_synthesis}}.

\begin{acknowledgement}
J.L.M.R. and T.P. wish to express their sincere gratitude to the Bavarian Ministry of Economic Affairs, Regional Development and Energy and TUMint. Energy Research GmbH for their generous support.  E.A. and E.P. acknowledge partial funding from the National Science Foundation DMREF Awards 1922090, 1922311, 1922372, the Office of Naval Research (ONR) under contract N00014-20-1-2280. 

\end{acknowledgement}


\bibliography{acs-achemso}

\newpage
\clearpage
\appendix

\setcounter{figure}{0}
\renewcommand{\thefigure}{A\arabic{figure}}
\setcounter{table}{0}

\renewcommand{\thetable}{A\arabic{table}}

\begin{center}

{\LARGE Supplementary Information for: Language Models Enable Data-Augmented Synthesis Planning for Inorganic Materials}

\vspace{0.5cm}

{\large
Thorben Prein$^{\dagger,\ddagger,\P}$, Elton Pan$^{\S}$, Janik Jehkul$^{\dagger}$, Steffen Weinmann$^{\dagger}$, Elsa A.\ Olivetti$^{\S}$, and Jennifer L.\ M.\ Rupp$^{*,\dagger,\P}$
}

\vspace{0.5cm}

$^{\dagger}$Technische Universität München, Garching b. München, Germany.\\
$^{\ddagger}$Munich Data Science Institute, Technische Universität München, Munich, Germany.\\
$^{\P}$TUMint. Energy Research GmbH, Munich, Germany.\\
$^{\S}$Department of Materials Science and Engineering, Massachusetts Institute of Technology, Cambridge, MA, USA.\\

\texttt{*}Corresponding author: \texttt{jrupp@tum.de}

\end{center}

\subsection{Dataset}
\label{appendix:precursor_suggestion}
All experiments were conducted using the dataset of Kononova \emph{et al.} \cite{kononova2019text}, which comprises 33\,343 text-mined solid-state synthesis recipes extracted via paragraph- and phrase-level NLP classifiers to identify targets, precursors, and byproducts. After removing entries with ambiguous stoichiometries and enforcing elemental consistency between targets and precursors, 18,804 reactions remain, of which 9,255 are unique. We adopt the year-based, Distinct Reactions, and Novel Material Systems (NMS) splits as defined by Prein \emph{et al.}~\cite{prein2025retro}.

\begin{figure}[htbp]
    \centering
    \includegraphics[width=0.8\textwidth]{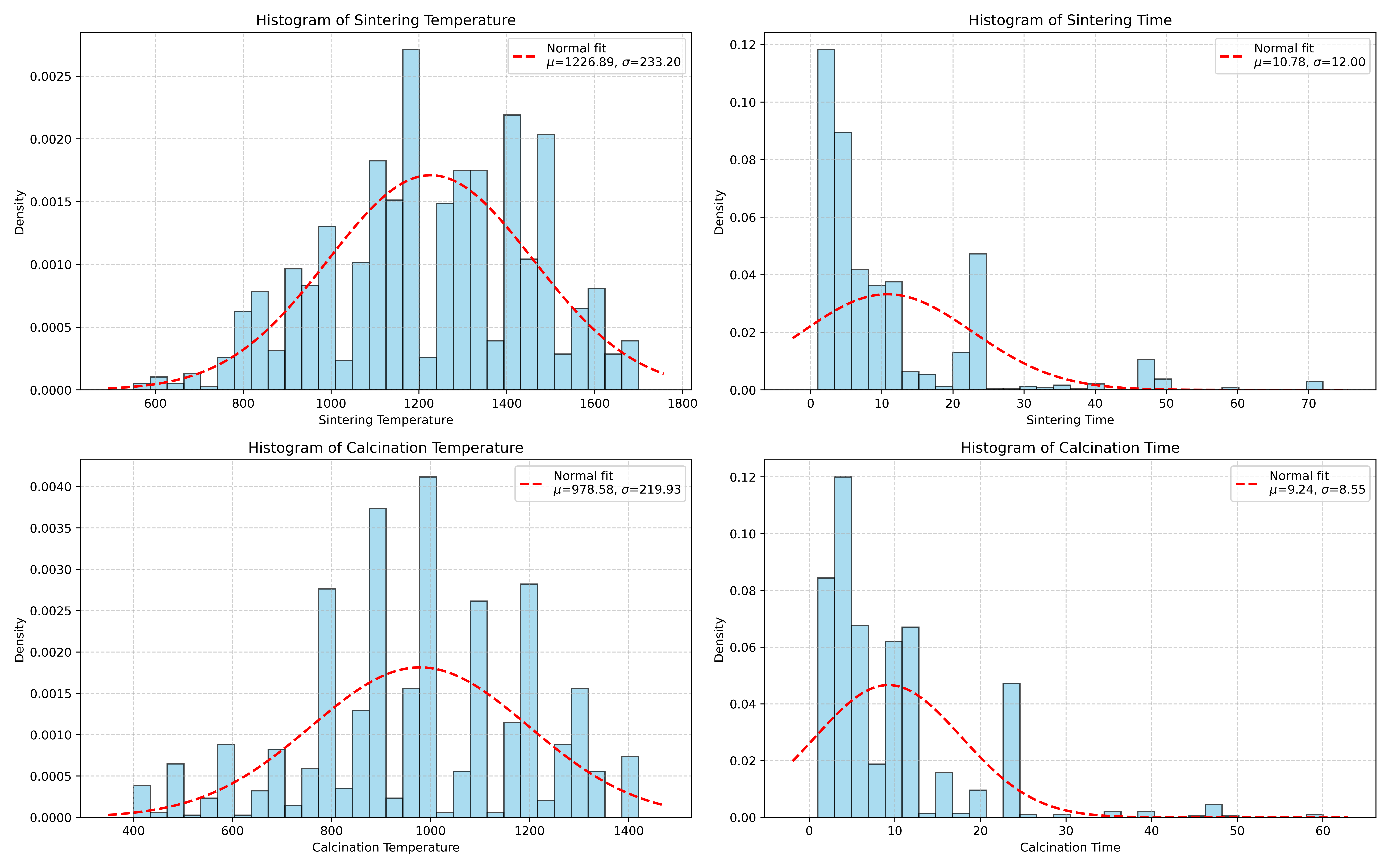} 
    \caption{Summary statistics of the regression subset of synthesis-condition data used for our evaluation.}
    \label{histograms}
\end{figure}

\begin{figure}[htbp]
    \centering
    \includegraphics[width=0.99\textwidth]{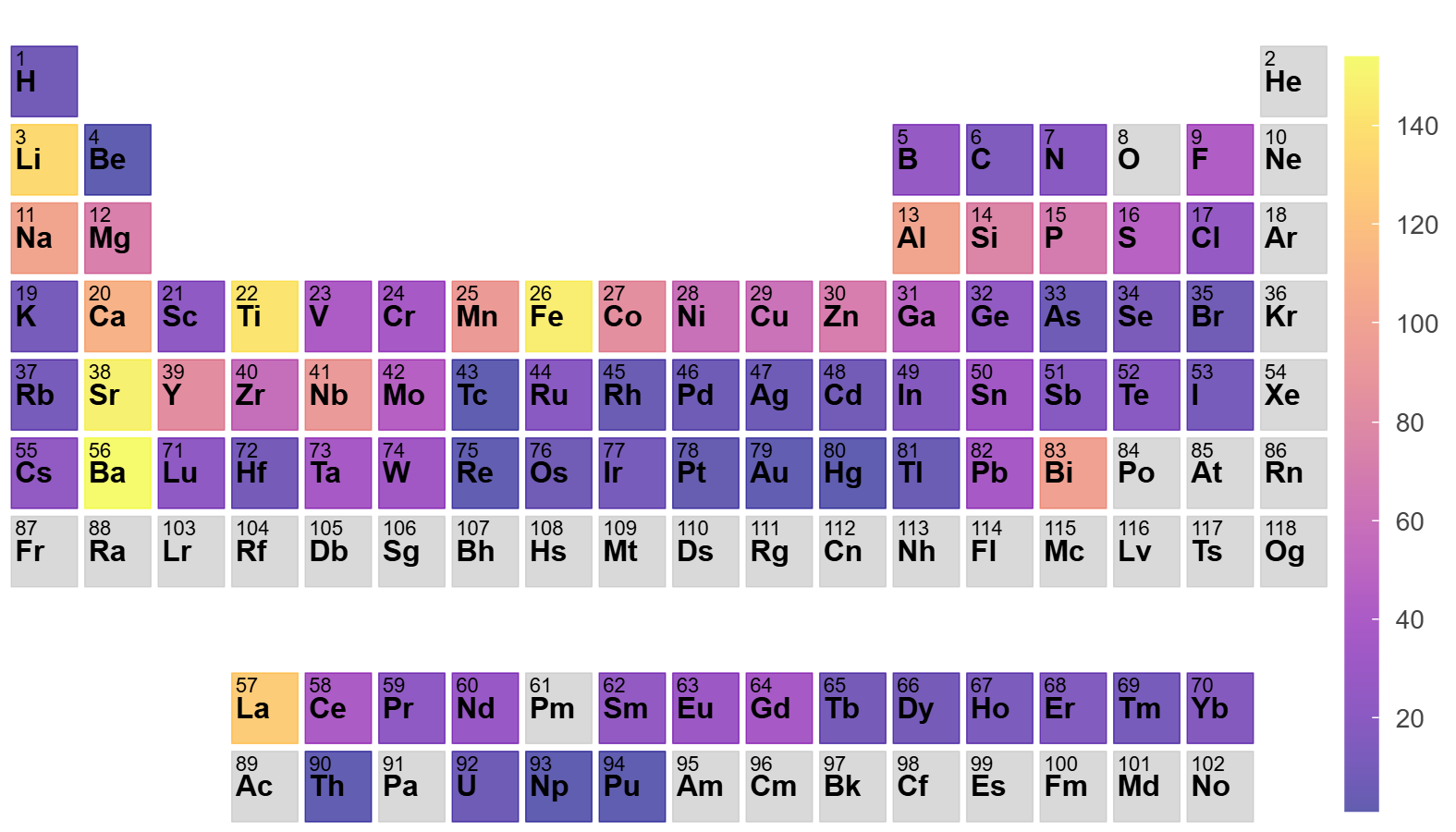} 
    \caption{Periodic table heatmap illustrating the frequency distribution of elements within the target material formulas (O excluded)}
\end{figure}

\subsection{LM Overview}\label{app:llm_outline}
In this study, we systematically evaluated seven contemporary LMs, selected to represent a diverse array of architectures, parameter scales and licensing schemes. The main characteristics of these models are summarized in Table~\ref{tab:llm_overview}.\\
\begin{table}[ht]
\centering
\resizebox{\textwidth}{!}{%
\begin{tabular}{@{}lllllllll@{}}
\toprule
\textbf{Model} & \textbf{Release} & \textbf{\#Params} & \textbf{\#Active} & \textbf{Context Window} & \textbf{Open Source} & \textbf{ELO Score} & \textbf{MMLU-Pro} & \textbf{Training Data} \\ \midrule
Qwen 2.5 VL-72B    & Jan 2025 & 72B   & -    & 32k tokens   & Yes (Apache 2.0)                & 1123 & 71.2 & 4T tokens \\
Mistral Small 3.1  & Mar 2025 & 24B   & -    & 128k tokens  & Yes (Apache 2.0)                & 1249 & 66.8 & -          \\
DeepSeek-V3-0324   & Mar 2025 & 671B  & 37B  & 164k tokens  & Yes (MIT License)               & 1369 & 81.2 & 14.8T tokens \\
Gemini 2.0 Flash   & Feb 2025 & -     & -    & 1M tokens    & No                              & 1352 & 76.4 & -          \\
GPT-4.1            & Apr 2025 & -     & -    & 1M tokens    & No                              & 1365 & -    & -          \\
LLaMA 4 Maverick   & Apr 2025 & 400B  & 17B  & 1M tokens    & Yes (LLaMA 4 Comm. License)     & 1266 & 80.5 & $\sim$22T tokens \\
Grok 3 Mini Beta   & Apr 2025 & -     & -    & 131k tokens  & No                              & -    & 78.9   & -          \\
\bottomrule
\end{tabular}%
}
\caption{LMs evaluated in this work, sorted by release date. Arena scores from the Chatbot Arena leaderboard as of May 18, 2025}
\label{tab:llm_overview}
\end{table}\\
\paragraph{Qwen 2.5 VL-72B.} Released in January 2025, Qwen 2.5 VL-72B (Apache 2.0 license) comprises 72 billion parameters and features a context window of 32,000 tokens.  The model achieved an ELO rating of 1123 and an MMLU-Pro score of 71.2. \cite{qwen-25-72b-vl, lmarena2025leaderboard}

\paragraph{Mistral Small 3.1.} This model was introduced in March 2025 under the Apache 2.0 license and contains 24 billion parameters. It offers an extended context window of 128,000 tokens. While detailed training data information remains undisclosed, it attains an ELO score of 1249 and an MMLU-Pro score of 66.8, indicating robust generalization abilities. \cite{mistrall-small-3.1, lmarena2025leaderboard}

\paragraph{DeepSeek-V3-0324.} This model, released under the MIT license, is a Mixture-of-Experts (MoE) language model featuring 671 billion total parameters, with 37 billion activated per token. Its MoE layers consist of 1 shared and 256 routed experts, with 8 routed experts actively engaged for each token. It's pre-trained on 14.8 trillion tokens and supports a context window extended to 128,000 tokens. The March 2025 checkpoint of this model achieves the highest ELO (1369) and MMLU-Pro (81.2) score among the evaluated models. \cite{deepseek-v3, deepseekv30324, lmarena2025leaderboard}

\paragraph{Gemini 2.0 Flash.} Released in February 2025 as a proprietary model, Gemini 2.0 Flash features a context window of 1 million tokens. While the specific parameter count and training details are not publicly disclosed, the model's performance is notable, reflected by an ELO rating of 1352 and an MMLU-Pro score of 76.4. \cite{gemini-2.0-flash, lmarena2025leaderboard}

\paragraph{GPT-4.1.} This model was introduced in April 2025, also supporting a large context window of 1 million tokens. Despite undisclosed training and parameter specifics, it achieved a remarkably high ELO score of 1365. \cite{gpt-4.1, lmarena2025leaderboard}

\paragraph{LLaMA 4 Maverick.} The model, released in April 2025 under the LLaMA 4 Community License, incorporates a Mixture-of-Experts (MoE) architecture with 128 experts. It features a grand total of 400 billion parameters, of which approximately 17 billion are actively engaged during inference. Trained on an extensive dataset of roughly 22 trillion tokens, its 1 million token context window supports advanced in-context learning capabilities. Evaluation metrics include an ELO score of 1266 and an MMLU-Pro score of 80.5.
 \cite{llama4-maverick, lmarena2025leaderboard}

\paragraph{Grok 3 Mini Beta.} This proprietary LM, launched in April 2025, provides a context window of 131,000 tokens. It attained an MMLU-Pro score of 78.9. The ELO score was not disclosed publicly. \cite{grok-3}

\subsection{Multi‑Provider Inference via OpenRouter Proxy}
\label{sec:openrouter_inference}

All model inference is performed through the OpenRouter API, which federates requests to multiple upstream providers under a single authentication and billing framework. For every model we set the temperature parameter at $\tau = 0.1$, giving token
probabilities
\begin{equation} \label{eq:softmax} p_\tau(i)=\frac{\exp(z_i/\tau)}{\sum_j \exp(z_j/\tau)} \end{equation}
and retained all other provider defaults. All models in Table \ref{tab:llm_overview} were benchmarked via OpenRouter on 1000 held‑out targets. These models were selected to represent a diverse cross-section of leading commercially available large language models, with a primary focus on balancing strong performance and cost-effectiveness.

\begin{table}[ht]
  \centering
  \scriptsize                   
  \begin{tabular*}{0.8\textwidth}{@{\extracolsep{\fill}} l c c }
    \toprule
    \textbf{Model} & \textbf{Input (\$/1M)} & \textbf{Output (\$/1M)} \\
    \midrule
    GPT-4.1 (OpenAI)               & 2.0  & 8.0  \\
    Grok 3 Mini Beta (xAI)         & 0.3  & 0.5  \\
    Llama 4 Maverick (Meta)        & 0.17 & 0.85 \\
    DeepSeek Chat v3 (DeepSeek)     & 0.27 & 1.1  \\
    Mistral Small 3.1 (Mistral)     & 0.10 & 0.30 \\
    Gemini 2.0 Flash-001 (Google)   & 0.10 & 0.40 \\
    Qwen 2.5 VL 72B (Alibaba)       & 0.70 & 0.70 \\
    \bottomrule
  \end{tabular*}
 \caption{Per-token API costs via OpenRouter (May 23, 2025)}
\end{table}

All API requests and responses are recorded in a structured log for full traceability. If a response doesn't satisfy the prescribed output format (missing the required list or dictionary structure), the identical prompt is retried up to two additional times, for a total of three format validation attempts. Should all three attempts yield structurally invalid output, the response is marked as a failure. API-level errors are immediately assigned a value of \texttt{None}. They are logged, and retried without counting against the three allowed format attempts. This separation ensures that transient infrastructure issues do not penalize model performance estimates.

\subsection{Prompting}
In-context learning refers to a language model’s ability to solve a novel task from prompt-side demonstrations without parameter updates. Performance in this setting typically improves with the number of examples (or ``shots'') provided, at least until the model’s context window is exhausted. This phenomenon was first highlighted in GPT-3, where accuracy on benchmarks such as SuperGLUE increased monotonically as the number of demonstrations was varied from 1--32, demonstrating that richer prompts can substitute for task-specific fine-tuning \cite{brown2020language}. Here, we adopt this methodology to determine an appropriate number of examples for our models. Our evaluation using Mistral Small on the precursor suggestion task shows that overall performance improves up to 40 in-context examples and plateaus beyond this point. Accordingly, we use 40 examples in our evaluations on all models.

\begin{figure}[htbp]
    \centering
    \includegraphics[width=0.99\textwidth]{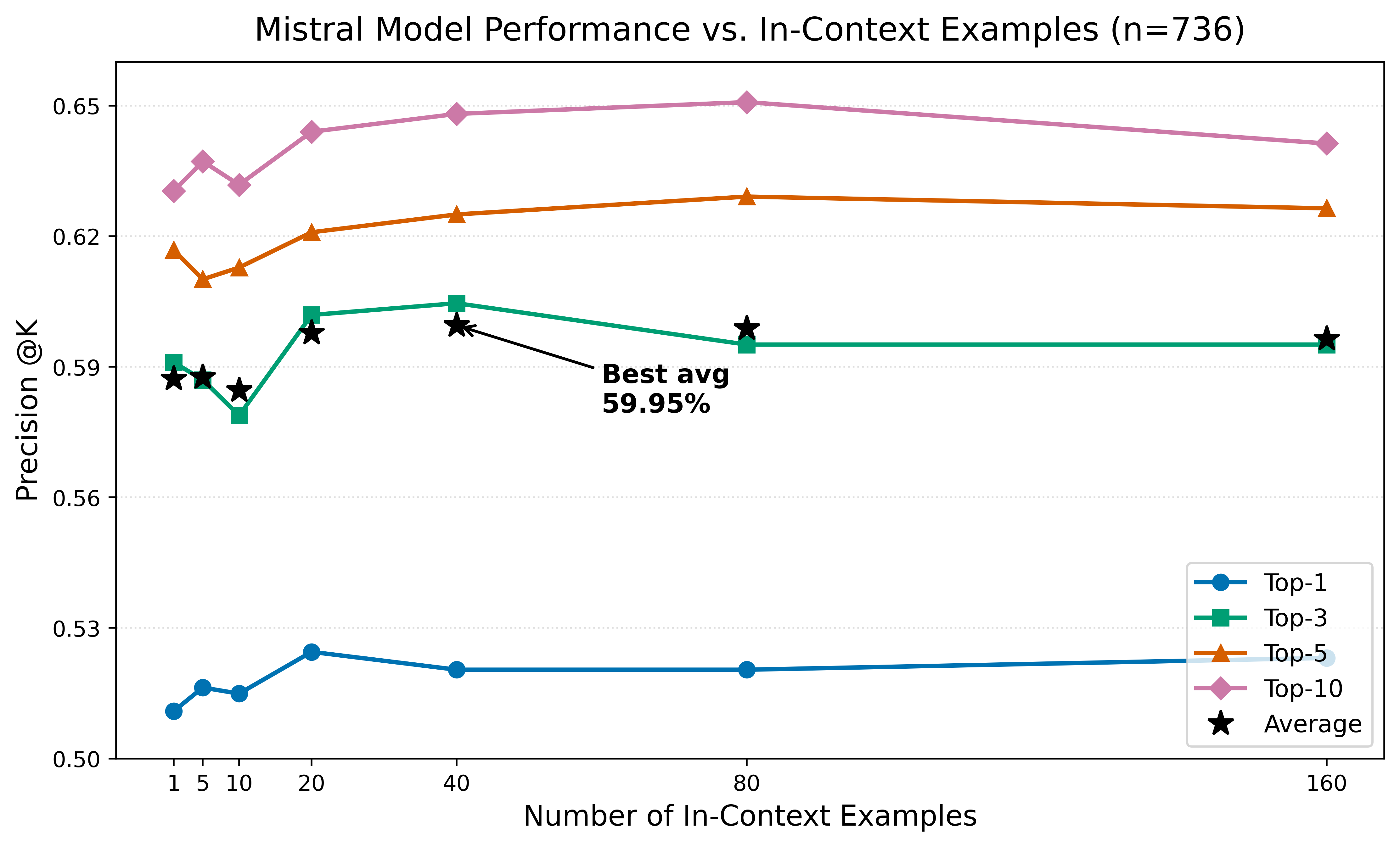} 
    \caption{Influence of the number of in-context examples on the performance across 736 samples of the validation set.}
    \label{fig:hyperparams-in-context}
\end{figure}

\subsection{Evaluation Protocols}
For the two main tasks at hand: Precursor suggestion and synthesis condition regression, we use the evaluation protocols described below. 

\subsubsection{Precursor Suggestion Task}
We use a subset of 1000 samples within the curated dataset from Kononova et al. \cite{kononova2019text}. In the precursor suggestion task, we prompt LMs to generate feasible precursors for a given target material. The model is provided with a subset of 40 in-context examples of solid-state synthesis to illustrate the required return format. It is instructed to output its answer as a Python object, which is subsequently parsed into a list of lists. The rank of each precursor set corresponds to its index within the list. We then parse all compounds and normalize them to a universal pretty formula format using \texttt{pymatgen}\cite{togo2024spglib}, before comparing the generated sets to the ground-truth precursor set. In the evaluation, we follow the protocol employed previously \cite{noh2024retrieval, prein2023mtencoder}. We iterate through the test dataset and assume a single ground-truth precursor set per generation and do not resolve for other reported ground-truth precursor sets regarding the same target material in our dataset, thereby reducing bias from frequently occurring entries in the dataset that an LM might retrieve easily. Note, that the dataset is cleaned from duplicate target material precursor pairs. We compute exact-match accuracy by counting the number of cases in which the ground-truth label appears among the top-1, top-3, etc. predictions.

\subsubsection{Synthesis-Condition Regression Task}

\begin{figure}[h!]
    \centering
    \includegraphics[width=0.6\textwidth]{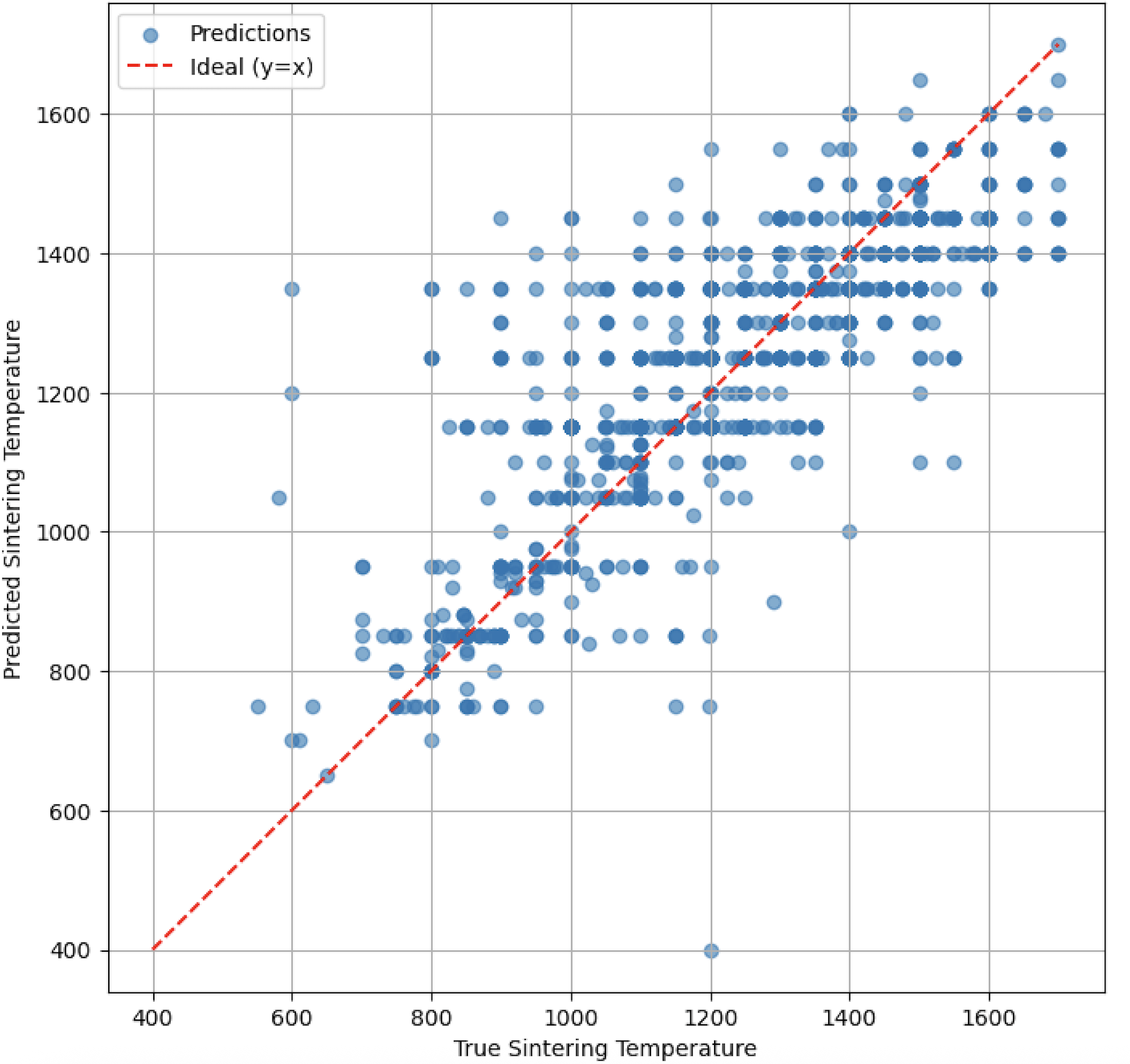} 
    \caption{Scatter plot of true versus predicted sintering temperatures for the Gemini 2.0 Flash model on the 40-shot test set with fixed precision length.}
    \label{fig:plot_results_conditions_gemini}
\end{figure}

For this task, we prompt the language model to predict the calcination and sintering temperatures together with their corresponding dwell times. We use a set of 1000 samples within the curated dataset from Kononova et al. \cite{kononova2019text}. Each prompt is prefaced with 40 few-shot examples drawn from the validation split (none of which appear in the test set).  After generation we extract the numeric values with a simple post-processor and compare them with the ground truth, reporting $R^2$, the mean absolute error (MAE), and the root-mean-square error (RMSE). Figure \ref{fig:plot_results_conditions_gemini} shows the results for the Gemini Flash model in the regression task for the sintering temperatures. Each scatter plot marks a data point.

\begin{table}[h!]
\centering
\caption{Mean predictor baseline: descriptive and error metrics for synthesis‐condition parameters.}
\label{tab:mean_predictor_baseline}
\begingroup
\footnotesize
\begin{tabular}{l c c c c c}
\toprule
\textbf{Parameter} 
  & \textbf{Mean} 
  & \textbf{Std.\ Dev.} 
  & \textbf{Norm.\ Std.\ (Std/Mean)} 
  & \textbf{MAE} 
  & \textbf{RMSE} \\
\midrule
Sintering temperature   
  & 1,230  & 233   & 0.190  & 192  & 233  \\ 
Sintering time          
  & 10.8    & 12.0  & 1.11   & 8.49 & 12.0 \\ 
Calcination temperature 
  & 979     & 220   & 0.225  & 177  & 220  \\ 
Calcination time        
  & 9.24    & 8.56  & 0.926  & 6.18 & 8.55 \\ 
\bottomrule
\end{tabular}
\endgroup
\end{table}

\subsection{Additional Dataset Generation Details}
In addition to the procedures described in the main text, we retrieved all 48,927 reported syntheses from the Materials Project database \cite{jain2013commentary}, filtering exclusively for compounds with documented experimental syntheses. Each chemical formula was parsed into a \texttt{Composition} object in \texttt{pymatgen} and converted into an atomic‐fraction vector over the union of all elements. These high‐dimensional vectors were discretized into \(M=1,000\) clusters via MiniBatchKMeans, producing a histogram representation of compositional regions. To select a uniform, maximally diverse subset of \(K=10,000\) formulas, we employed a greedy Shannon‐entropy maximization algorithm: at each iteration, a random candidate subset was sampled, the increase in histogram entropy for each candidate was computed, and the candidate yielding the greatest gain was selected. We then prompted GPT-4.1 to assess solid‐state synthesizability, flagged compositions were excluded and only those amenable to solid-state methods were retained. For each remaining material, the top three precursor routes, reflecting the model’s 64.1 \% Top-3 accuracy, were preserved (Table~\ref{tab:model_performance_llm_precursor}). Synthesis-condition parameters were subsequently predicted by GPT-4.1 (Table~\ref{tab:temp_performance}), yielding 29,473 entries. The initial review revealed unphysically low temperatures arising from liquid-phase route suggestions (e.g., \ch{Cu(NH3)4(NO3)2} via crystallization rather than solid-state synthesis \cite{morosin1976crystal}). To ensure plausibility, we imposed minimum temperature thresholds of 300\,\textdegree{}C for calcination and 500\,\textdegree{}C for sintering, resulting in the final set of 28,548 solid-state synthesis recipes.

\begin{table}[h!]
\centering
\small 
\caption{Example entries from the generated synthesis dataset. Each row shows precursor combinations and synthesis conditions for a target material.}
\begin{tabular}{>{\centering\arraybackslash}m{4.2cm}
                >{\centering\arraybackslash}m{2.5cm}
                >{\centering\arraybackslash}m{1.2cm}
                >{\centering\arraybackslash}m{1.2cm}
                >{\centering\arraybackslash}m{1.2cm}
                >{\centering\arraybackslash}m{1.2cm}}
\toprule
\textbf{Precursor Compounds} & \textbf{Target Compound} & \textbf{Sinter Temp. (°C)} & \textbf{Time (h)} & \textbf{Calc Temp. (°C)} & \textbf{Time (h)} \\
\midrule
\texttt{\{Cs$_2$CO$_3$, SiO$_2$, H$_2$O\}} & CsSi$_3$HO$_7$ & 950 & 6 & 700 & 6 \\
\texttt{\{NaOH, CO$_2$\}} & Na$_2$CO$_3$ & 350 & 2 & 200 & 2 \\
\texttt{\{PbS, As$_2$S$_3$, Pb\}} & As$_4$(Pb$_3$S$_5$)$_3$ & 700 & 12 & 500 & 6 \\
\texttt{\{Tb$_4$O$_7$, SeO$_2$\}} & Tb$_2$Se$_2$O$_7$ & 900 & 6 & 650 & 6 \\
\texttt{\{CoCl$_2$, B$_{10}$H$_{14}$, C$_6$H$_6$\}} & CoB$_{10}$H$_{21}$(C$_6$Cl)$_2$ & 350 & 6 & 250 & 8 \\
\texttt{\{BaCO$_3$, Fe$_2$O$_3$, GeO$_2$\}} & Ba$_2$FeGe$_2$O$_7$ & 1250 & 6 & 1000 & 4 \\
\texttt{\{MoO$_3$, C$_2$N$_2$O, CO$_2$\}} & Mo$_2$C$_5$N$_3$O$_{14}$ & 900 & 6 & 750 & 6 \\
\texttt{\{Li$_2$CO$_3$, In$_2$O$_3$, Ir\}} & Li$_2$InIr & 1100 & 12 & 700 & 6 \\
\texttt{\{Y$_2$O$_3$, Al$_2$O$_3$, Co$_3$O$_4$\}} & YAl$_2$Co & 1450 & 24 & 1200 & 10 \\
\texttt{\{Ag$_2$CO$_3$, TeO$_2$, MoO$_3$\}} & Ag$_2$Te$_4$MoO$_{12}$ & 650 & 6 & 500 & 6 \\
\bottomrule
\end{tabular}
\label{tab:representative_samples}
\end{table}

\newpage
\subsection{SyntMTE Ablations}

\begin{figure}[htbp]
    \centering
    \caption{Parity plots for the regression task shown in Table \ref{tab:synthetic_data_augmentation}. The model in a) has been trained on literature and synthetic data, while the model in b) is fine-tuned on synthetic data only. Notably $R^2$ scores are increased in the second setting, while MAEs increase.}
    \label{fig:parity_r2}
    \includegraphics[width=0.7\textwidth]{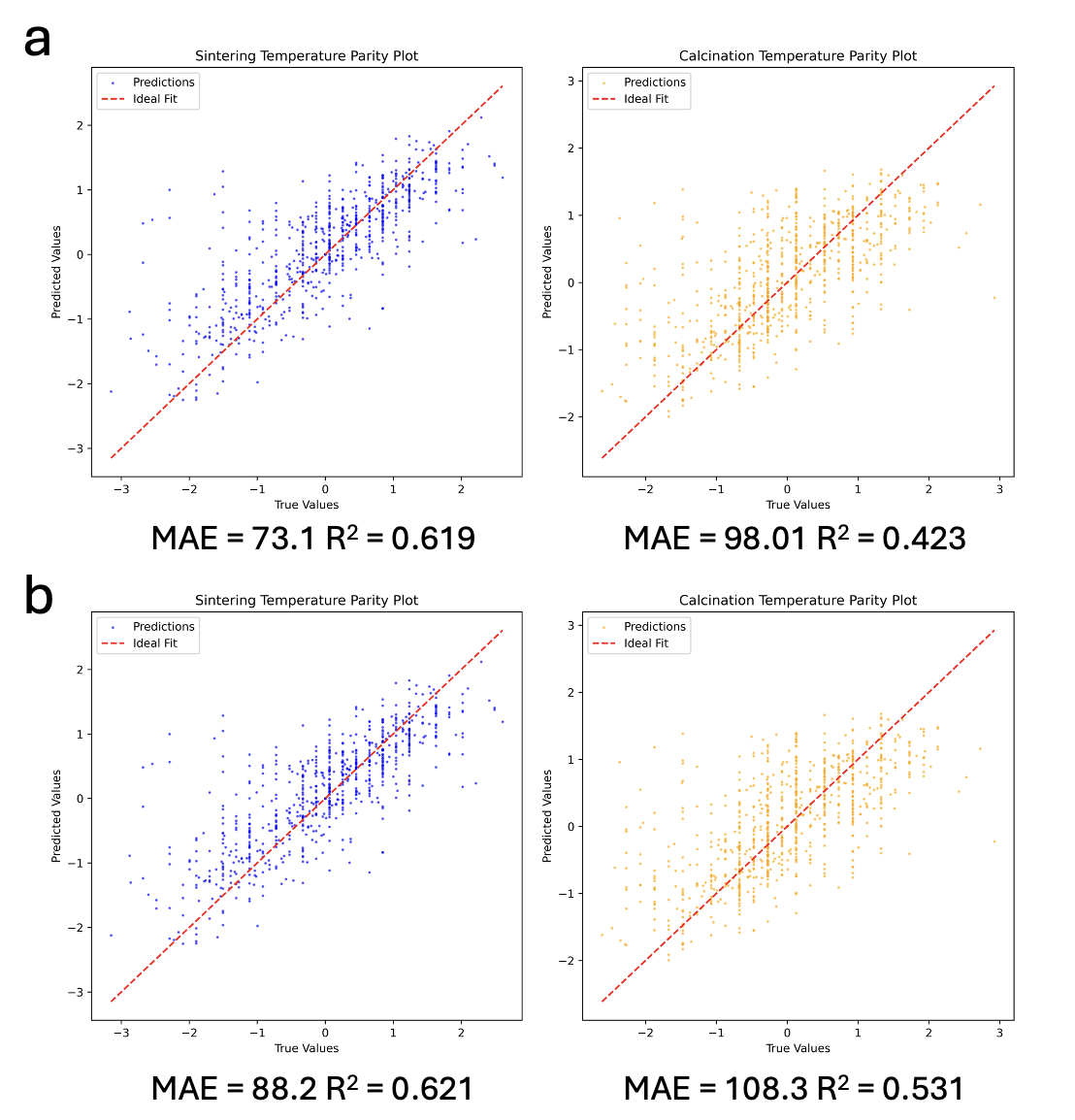}
\end{figure}

We systematically replaced \texttt{SyntMTE}’s mean-pool aggregator with eight alternatives ranging from parameter-free statistics to fully learned sequence encoders (Table~\ref{tab:ablation_pooling}, Fig.~\ref{fig:parity_r2}) using an identical literature-stage only fine-tuning schedule to isolate pooling as the sole variable. Averaging precursor and product embeddings yields the lowest absolute errors: 75.5\,°C for sintering and 103.3\,°C for calcination and matches the plain sum operator in producing the highest coefficient of determination for sintering, indicating that MTEncoder already encodes salient chemistry and that additional learnable mixing adds variance without signal improvement. Replacing statistical pooling with an LSTM, or self-attention block only marginally improves $R^2$ for calcination and leaves sintering performance unchanged or degraded, with large standard errors (Table~\ref{tab:ablation_pooling}) suggesting no benefit from the additional parameters. Given its robustness, lack of additional parameters, and consistently superior mean absolute error, we adopt mean pooling as the default aggregation scheme for all subsequent \texttt{SyntMTE} experiments.  

Figure~\ref{fig:parity_r2} shows parity plots of predicted versus experimental temperatures under two fine‐tuning regimes: (a) combined synthetic\,+\,literature data and (b) synthetic data only pre-training. In regime (a), predictions closely follow measured values across the entire temperature range, with mean absolute errors of 73\,$^\circ$C (sintering) and 98\,$^\circ$C (calcination), preserving accuracy at both central and extreme values. By contrast, regime~(b) reduces variance near the mean, increasing the coefficient of determination to $R^2=0.620$ (sintering) and 0.53 (calcination), but systematically underestimates the highest and lowest temperatures. These results indicate that synthetic-only pre-training improves overall fit by mitigating outliers but compromises the accurate representation of empirical long-tail behavior essential for modeling real-world processing.

\begin{table}[h!]
  \centering
  \caption{%
    \textbf{Test Set Performance Across Pooling Modes.}
    Mean ± SD across 5 runs in parentheses.
  }
  \label{tab:ablation_pooling}
  \renewcommand{\arraystretch}{0.8}
  \tiny                              
  \begin{adjustbox}{width=0.8\textwidth,center}
    \begin{tabular}{@{}l cc cc@{}}
      \toprule
      & \multicolumn{2}{c}{\textbf{Sintering Temp}} 
      & \multicolumn{2}{c}{\textbf{Calcination Temp}} \\
      \cmidrule(lr){2-3}\cmidrule(lr){4-5}
      \textbf{Mode}
        & \textbf{MAE↓} & \textbf{R²↑}
        & \textbf{MAE↓} & \textbf{R²↑} \\
      \midrule
      concat    & \makecell{81.40\\(1.54)} & \makecell{0.542\\(0.010)}
                & \makecell{109.0\\(2.55)} & \makecell{0.320\\(0.032)} \\
      precur    & \makecell{76.61\\(2.10)} & \makecell{0.584\\(0.015)}
                & \makecell{104.5\\(2.30)} & \makecell{0.355\\(0.026)} \\
      lstm      & \makecell{79.02\\(4.66)} & \makecell{0.569\\(0.033)}
                & \makecell{\underline{103.6}\\(2.18)} & \makecell{\underline{0.374}\\(0.012)} \\
      conv      & \makecell{77.77\\(1.14)} & \makecell{0.579\\(0.012)}
                & \makecell{106.1\\(1.12)} & \makecell{0.341\\(0.017)} \\
      mean      & \makecell{\textbf{75.51}\\(1.07)} & \makecell{\underline{0.598}\\(0.007)}
                & \makecell{\textbf{103.3}\\(0.84)} & \makecell{0.371\\(0.014)} \\
      mean\_m   & \makecell{79.68\\(2.31)} & \makecell{0.569\\(0.018)}
                & \makecell{104.8\\(2.06)} & \makecell{0.358\\(0.015)} \\
      sum       & \makecell{\underline{75.68}\\(1.05)} & \makecell{\textbf{0.599}\\(0.012)}
                & \makecell{103.8\\(1.63)} & \makecell{0.363\\(0.018)} \\
      max       & \makecell{77.89\\(3.27)} & \makecell{0.573\\(0.027)}
                & \makecell{104.5\\(2.44)} & \makecell{0.352\\(0.024)} \\
      attention & \makecell{76.66\\(0.83)} & \makecell{0.585\\(0.008)}
                & \makecell{103.95\\(1.24)} & \makecell{\textbf{0.378}\\(0.012)} \\
      \bottomrule
    \end{tabular}
  \end{adjustbox}
\end{table}

\newpage

\subsection{Baseline Models}
For the regression task we employ serveral baseline models.

\subsection{CrabNet}
CrabNet is a composition-only materials prediction framework that leverages a transformer encoder to learn contextualized embeddings for each element in a compound \cite{wang2021compositionally}. By combining learned element vectors with sinusoidal “fractional embeddings” of stoichiometry, CrabNet’s multi-head self-attention layers capture complex inter-element interactions without any hand-crafted descriptors or structural information. This design not only yields state-of-the-art accuracy on benchmarks like MatBench, often outperforming graph-based models such as Roost, \cite{goodall2020predicting} but also enables interpretability: attention maps highlight which element pairs drive a given property prediction. In our experiments we use three transformer blocks.

\subsection{MTEncoder (\texttt{SyntMTE})}
\label{appendix_mte}
\begin{figure}[htb!]
    \centering
    \includegraphics[width=0.35\columnwidth]{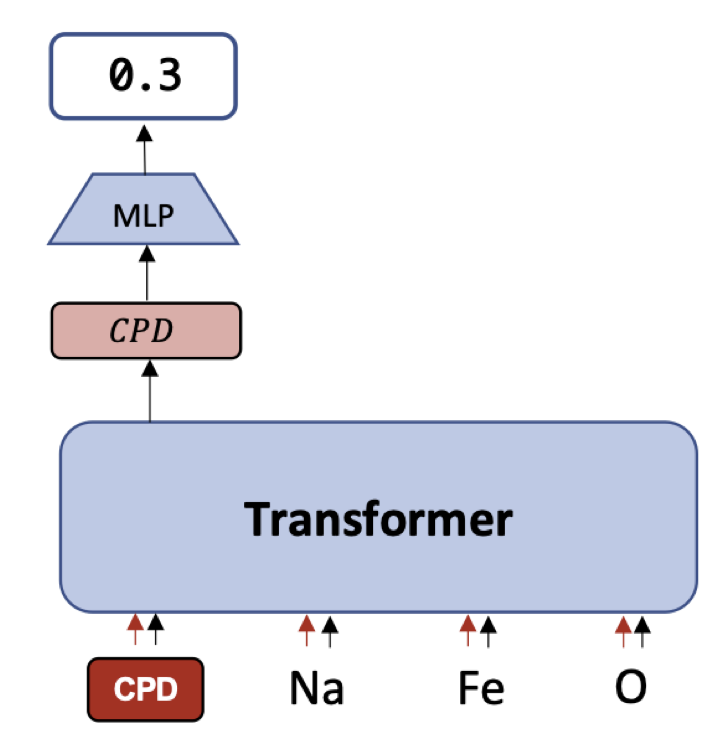}
    \caption{\textbf{Overview of the MTEncoder pipeline.}
    Material compositions are tokenized and processed by a transformer to generate feature embeddings for downstream property prediction.}
    \label{fig:plotmte}
\end{figure}

Figure~\ref{fig:plotmte} depicts the MTEncoder workflow, demonstrating how a material’s elemental representation is encoded via a transformer-based model \cite{prein2023mtencoder}. Each material is broken down into individual element tokens (e.g., Na, Fe, O) alongside a dedicated “Compound” token (\textit{CPD}) that aggregates the element-specific information. These tokens are fed into the transformer encoder, which produces context-rich embeddings for the composition. The embedding associated with the CPD token serves as the learned representation of the material and is passed to an MLP head to predict various properties. Pretraining is conducted using the Alexandria database on 12 tasks (Table \ref{tab:pretraining_tasks}\cite{schmidt2024improving}).

\begin{table}[h]
    \centering
    \begin{tabular}{l}
        \toprule
        \textbf{Pretraining Objectives} \\
        \midrule
        Stress \\
        Band Gap (Direct) \\
        Band Gap (Indirect) \\
        Density of States at Fermi Level \\
        Energy Above Hull \\
        Formation Energy \\
        Corrected Total Energy \\
        Phase Separation Energy \\
        Number of Atomic Sites \\
        Total Magnetic Moment \\
        Crystal Space Group \\
        Masked Element Reconstruction (Self-Supervised) \\
        \bottomrule
    \end{tabular}
    \caption{\textbf{Pretraining objectives for MTEncoder.}
    These tasks are drawn from the Alexandria materials dataset \cite{schmidt2024improving}.}
    \label{tab:pretraining_tasks}
\end{table}

\subsection{Composition + NN}
Three-layer feed-forward neural network: This model comprises three fully connected layers, each followed by a ReLU activation, optional dropout for regularization, and layer normalization to accelerate convergence and stabilize training. Compositional feature vectors (for example, elemental fractional embeddings) are input to this multilayer perceptron (MLP) to predict the target property.

\subsection{Composition + XGB}
A gradient-boosted decision-tree model trained on the same compositional features. XGBoost captures nonlinear interactions among elemental descriptors through an ensemble of shallow trees, providing an efficient and robust baseline for materials-property prediction \cite{chen2016xgboost}.

\newpage
\newpage
\newpage
\newpage
\newpage
\newpage
\newpage
\newpage

\subsection{Prompt Template}

Listing~\ref{lst:prompt_na3bi} shows the full prompt used for predicting precursor combinations for the solid-state synthesis of \ce{Na3Bi(AsO4)2}. The prompt explicitly instructs the model to generate 20 plausible precursor combinations suitable for solid-state synthesis, ensuring that each combination contains all required elements from the target material, favors stable and common laboratory reagents and ranks combinations by chemical plausibility.

\begin{lstlisting}[language=Python, caption={Prompt for precursor generation}, label={lst:prompt_na3bi}]
You are a computational chemistry expert specializing in solid-state synthesis and retrosynthesis.
Your task is to identify potential precursor combinations for solid-state synthesizing the target material: 'Na3Bi(AsO4)2'.

**Requirements**:
1.  Generate 20 distinct combinations of precursor materials.
2.  Use standard chemical formulas ONLY (e.g., 'TiO2', 'Na2CO3').
3.  **Constraint Check:** Ensure each precursor combination contains ALL elements present in the target material 'Na3Bi(AsO4)2'. Assume Oxygen and other common laboratory elements (e.g., C for carbonate sources) are available.
4.  **Plausibility Filter:** Prefer chemically plausible routes using reasonably common and stable laboratory reagents. A plausible route is one that uses precursors commonly found in solid-state synthesis and avoids highly unstable or rare compounds.
5.  Order the 20 combinations from the MOST plausible/common synthesis routes to the LEAST plausible/common.
6.  **Common Precursor Types:** Consider oxides (e.g., TiO2, Fe2O3), carbonates (e.g., Na2CO3, CaCO3), nitrates (e.g., KNO3, Ca(NO3)2), hydroxides (e.g., Al(OH)3), and other standard laboratory reagents.
7.  **No Gases:** Do not include 'O2' in the precursor combinations.
7.  If the target material is not suitable for solid-state synthesis, respond with False as a boolean.

**Examples of Target -> Precursors:**
- Target: "\ce{MoF5}", Precursors: [False] #only synthesizable via gas-solid reaction, not suitable for conventional solid-state synthesis
- Target: 'Gd2TiO5', Precursors: ['TiO2', 'Gd2O3']
- Target: 'NdTl(MoO4)2', Precursors: ['MoO3', 'Tl2O3', 'Nd2O3']
- Target: 'Sr(GaO2)2', Precursors: ['SrCO3', 'Ga2O3']
- Target: 'La0.075Ta2O5.113', Precursors: ['La2O3', 'Ta2O5']
- Target: 'LaFeO3', Precursors: ['Fe2O3', 'LaCO3']
- Target: 'Sr1.9Ca1Tl0.9V0.1Cu2Bi0.1O7', Precursors: ['SrCO3', 'Tl2O3', 'CuO', 'CaO', 'Bi2O3', 'V2O5']
- Target: 'La3RuO7', Precursors: ['RuO2', 'La2O3']
- Target: 'Zr0.8Ti1Sn0.2O4', Precursors: ['SnO2', 'TiO2', 'ZrO2']
- Target: 'La1Fe0.95W0.05O3', Precursors: ['WO3', 'Fe2O3', 'La2O3']
- Target: 'Bi3PO7', Precursors: ['PH9(NO2)2', 'Bi2O3']
- Target: 'CsTaWO6', Precursors: ['WO3', 'Cs2CO3', 'Ta2O5']
- Target: 'TaWO6', Precursors: ['WO3', 'Ta2O5']
- Target: 'Dy0.05Zn1Ga1.95O4', Precursors: ['Dy2O3', 'Ga2O3', 'ZnO']
- Target: 'Nd0.02Gd0.98V1O4', Precursors: ['Gd2O3', 'V2O5', 'Nd2O3']
- Target: 'Ba1Pr0.8In0.2O3', Precursors: ['BaCO3', 'Pr6O11', 'In2O3']
- Target: 'Ba4SrSmTi3V7O30', Precursors: ['SrCO3', 'Sm2O3', 'BaCO3', 'TiO2', 'V2O5']
- Target: 'ZrSiO', Precursors: ['SiO2', 'ZrO2']
- Target: 'Ba0.6Sr0.4Nb0.1Co0.9O3', Precursors: ['SrCO3', 'Nb2O5', 'BaCO3', 'Co2O3']
- Target: 'CsAlP2O7', Precursors: ['Cs2O', 'P2O5', 'Al2O3']
- Target: 'Ag4.64Pb2O5.87', Precursors: ['Ag2O', 'PbO']
- Target: 'Ca1Ti4Cu3.2O12', Precursors: ['CuO', 'TiO2', 'CaCO3']
- Target: 'Gd0.3Fe1Bi0.7O3', Precursors: ['Gd2O3', 'Bi2O3', 'Fe2O3']
- Target: 'YTiO', Precursors: ['TiO2', 'Y2O3']
- Target: 'Ba6Sn6Se13', Precursors: ['BaSe', 'Se', 'Sn']
- Target: 'V0.9Cu0.1Bi2O5.35', Precursors: ['CuO', 'Bi2O3', 'V2O5']
- Target: 'MgNb2(PbO3)3', Precursors: ['Nb2O5', 'MgCO3', 'PbO']
- Target: 'Cs0.75Rb0.25P1H2O4', Precursors: ['RbP(HO2)2', 'CsP(HO2)2']
- Target: 'TiCdO3', Precursors: ['TiO2', 'CdO']
- Target: 'Pu0.9Am0.1O2', Precursors: ['AmO2', 'PuO2']
- Target: 'La0.9Mn1Pb0.1O3', Precursors: ['PbO', 'MnO2', 'La2O3']
- Target: 'Li3.55Ca5.45Si3O12.45F1.55', Precursors: ['SiO2', 'CaCO3', 'Li2CO3', 'LiF']
- Target: 'HoMnO3', Precursors: ['Mn2O3', 'Ho2O3']
- Target: 'SiPbC', Precursors: ['Pb', 'SiC']
- Target: 'CsAl(SiO3)2', Precursors: ['SiO2', 'Cs2CO3', 'Al2O3']
- Target: 'CdIn2O4', Precursors: ['CdO', 'In2O3']
- Target: 'CdWO4', Precursors: ['WO3', 'CdO']
- Target: 'Sr1.8Ca0.9Ti0.2Tl0.9Cu2Bi0.1O7', Precursors: ['SrCO3', 'Tl2O3', 'Ti2O3', 'CuO', 'CaO', 'Bi2O3']
- Target: 'Ca3ZrSi2O9', Precursors: ['SiO2', 'CaCO3', 'ZrO2']
- Target: 'Ba3NbFe3(SiO7)2', Precursors: ['SiO2', 'Nb2O3', 'BaCO3', 'Fe2O3']
- Target: 'Mn2Ni(CO2)6', Precursors: ['MnH6(CO)4', 'NiH6(CO)4']

Note: Ensure the quality and consistency of the example data to prevent parsing issues.

Generate the list of 20 precursor combinations for the target 'Na3Bi(AsO4)2'.
**Output Format:** Respond ONLY with a single Python-formatted list of lists. Each inner list should contain the precursor strings. Typically, 2-4 precursors per combination are expected.
Example Output Format: [['precursor1a', 'precursor1b', 'precursor1c'], ['precursor2a', 'precursor2b', 'precursor2c'], , ..., ['precursor20a', 'precursor20b', 'precursor20c']]

**Important:** Ensure all combinations are chemically valid and contain all elements needed to synthesize the target material. Do not include any explanations or text outside the Python list format.
\end{lstlisting}

Conditions Prompt
\begin{lstlisting}
\begin{lstlisting}
 You are a computational chemistry expert specializing in solid-state synthesis.
Assume there is only one sintering and one calcination step involved.
Your task is to predict the optimal synthesis conditions for the following chemical reaction:
4 BaCO3 + 1 Fe2O3 + 4 Nb2O5 + 0.333 Pr6O11 == 1 Ba4Pr2Fe2Nb8O30 + 4 CO2 + 0.333 O2

**Required Conditions to Predict**:
1. Sintering Temperature (in \(^{\circ}\)C)
2. Sintering Time (in hours)
3. Calcination Temperature (in \(^{\circ}\)C)
4. Calcination Time (in hours)

**Guidelines for Prediction**:
- Base your predictions on established solid-state chemistry principles.
- Provide scientifically plausible values within typical laboratory ranges.
- Assume there is only one sintering and one calcination step involved.

**Examples of Synthesis Conditions:**
- Reaction: 2 BaCO3 + 0.667 Co3O4 + 6 Fe2O3 == 1 Ba2Co2Fe12O22 + 2 CO2 + 0.333 O2
  Sintering Temperature (\(^{\circ}\)C): 1240.0
  Sintering Time (hours): 4.0
  Calcination Temperature (\(^{\circ}\)C): 1000.0
  Calcination Time (hours): 6.0

- Reaction: 1 GeO2 + 1 ZnO == 1 ZnGeO3
  Sintering Temperature (\(^{\circ}\)C): 1240.0
  Sintering Time (hours): 6.0
  Calcination Temperature (\(^{\circ}\)C): 1000.0
  Calcination Time (hours): 6.0

- Reaction: 0.5 In2O3 + 0.5 La2O3 == 1 LaInO3
  Sintering Temperature (\(^{\circ}\)C): 1150.0
  Sintering Time (hours): 24.0
  Calcination Temperature (\(^{\circ}\)C): 830.0
  Calcination Time (hours): 1.0

- Reaction: 0.05 Fe2O3 + 0.015 O2 + 0.98 SrCO3 + 0.92 TiO2 == 1 Sr0.98Ti0.92Fe0.1O3 + 0.98 CO2
  Sintering Temperature (\(^{\circ}\)C): 1450.0
  Sintering Time (hours): 24.0
  Calcination Temperature (\(^{\circ}\)C): 1100.0
  Calcination Time (hours): 12.0

- Reaction: 1.5 Li2CO3 + 0.5 Nb2O5 == 1 Li3NbO4 + 1.5 CO2
  Sintering Temperature (\(^{\circ}\)C): 970.0
  Sintering Time (hours): 2.0
  Calcination Temperature (\(^{\circ}\)C): 800.0
  Calcination Time (hours): 4.0

- Reaction: 0.05 Al2O3 + 0.9 CaCO3 + 0.05 Nd2O3 + 0.9 TiO2 == 1 Ca0.9Nd0.1Ti0.9Al0.1O3 + 0.9 CO2
  Sintering Temperature (\(^{\circ}\)C): 1300.0
  Sintering Time (hours): 12.0
  Calcination Temperature (\(^{\circ}\)C): 1200.0
  Calcination Time (hours): 10.0

- Reaction: 1 BaCO3 + 1 Ir + 1 O2 == 1 BaIrO3 + 1 CO2
  Sintering Temperature (\(^{\circ}\)C): 1000.0
  Sintering Time (hours): 72.0
  Calcination Temperature (\(^{\circ}\)C): 900.0
  Calcination Time (hours): 12.0

- Reaction: 2 BaCO3 + 3 CuO + 0.5 Y2O3 == 1 Y1Ba2Cu3O + 2 CO2 + 2.75 O2
  Sintering Temperature (\(^{\circ}\)C): 950.0
  Sintering Time (hours): 8.0
  Calcination Temperature (\(^{\circ}\)C): 950.0
  Calcination Time (hours): 24.0

- Reaction: 1.5 Bi2O3 + 0.5 O2 + 0.5 Sb2O3 == 1 Bi3SbO7
  Sintering Temperature (\(^{\circ}\)C): 860.0
  Sintering Time (hours): 2.0
  Calcination Temperature (\(^{\circ}\)C): 700.0
  Calcination Time (hours): 4.0

- Reaction: 0.667 Co3O4 + 1 SnO2 == 1 Co2SnO4 + 0.333 O2
  Sintering Temperature (\(^{\circ}\)C): 1250.0
  Sintering Time (hours): 48.0
  Calcination Temperature (\(^{\circ}\)C): 1200.0
  Calcination Time (hours): 48.0

- Reaction: 0.5 O2 + 1 PbO + 2 TiO2 == 1 Pb(TiO3)2
  Sintering Temperature (\(^{\circ}\)C): 1100.0
  Sintering Time (hours): 2.0
  Calcination Temperature (\(^{\circ}\)C): 800.0
  Calcination Time (hours): 2.0

- Reaction: 0.8 CaCO3 + 0.2 SrCO3 + 1 TiO2 == 1 Ca0.8Sr0.2TiO3 + 1 CO2
  Sintering Temperature (\(^{\circ}\)C): 1460.0
  Sintering Time (hours): 4.0
  Calcination Temperature (\(^{\circ}\)C): 1100.0
  Calcination Time (hours): 4.0

- Reaction: 1 Li2CO3 + 1 TiO2 == 1 Li2TiO3 + 1 CO2
  Sintering Temperature (\(^{\circ}\)C): 1100.0
  Sintering Time (hours): 2.0
  Calcination Temperature (\(^{\circ}\)C): 600.0
  Calcination Time (hours): 2.0

- Reaction: 2 TiO2 + 1 Y2O3 == 1 Y2Ti2O7
  Sintering Temperature (\(^{\circ}\)C): 1600.0
  Sintering Time (hours): 6.0
  Calcination Temperature (\(^{\circ}\)C): 1000.0
  Calcination Time (hours): 6.0

- Reaction: 1 La2O3 + 2 ZrO2 == 1 La2Zr2O7
  Sintering Temperature (\(^{\circ}\)C): 1549.85
  Sintering Time (hours): 50.0
  Calcination Temperature (\(^{\circ}\)C): 1399.85
  Calcination Time (hours): 10.0

- Reaction: 0.5 Dy2O3 + 0.25 O2 + 1 RuO2 + 2 SrCO3 == 1 Sr2DyRuO6 + 2 CO2
  Sintering Temperature (\(^{\circ}\)C): 1199.85
  Sintering Time (hours): 24.0
  Calcination Temperature (\(^{\circ}\)C): 979.85
  Calcination Time (hours): 12.0

- Reaction: 0.025 Gd2O3 + 0.013 O2 + 0.95 SnO2 == 1 Gd0.05Sn0.95O2
  Sintering Temperature (\(^{\circ}\)C): 1350.0
  Sintering Time (hours): 24.0
  Calcination Temperature (\(^{\circ}\)C): 1200.0
  Calcination Time (hours): 24.0

- Reaction: 1 MgO + 1 TiO2 == 1 MgTiO3
  Sintering Temperature (\(^{\circ}\)C): 1400.0
  Sintering Time (hours): 4.0
  Calcination Temperature (\(^{\circ}\)C): 1100.0
  Calcination Time (hours): 4.0

- Reaction: 1.7 CaCO3 + 2 CuO + 0.15 La2O3 + 4 TiO2 == 1 Ca1.7La0.3Cu2Ti4O12 + 1.7 CO2 + 0.075 O2
  Sintering Temperature (\(^{\circ}\)C): 1100.0
  Sintering Time (hours): 24.0
  Calcination Temperature (\(^{\circ}\)C): 900.0
  Calcination Time (hours): 5.0

- Reaction: 0.05 Fe2O3 + 0.035 O2 + 0.98 SrCO3 + 0.9 TiO2 == 1 Sr0.98Ti0.9Fe0.1O3 + 0.98 CO2
  Sintering Temperature (\(^{\circ}\)C): 1450.0
  Sintering Time (hours): 24.0
  Calcination Temperature (\(^{\circ}\)C): 1100.0
  Calcination Time (hours): 12.0

- Reaction: 1 MoO3 + 1 Nd2O3 == 1 Nd2MoO6
  Sintering Temperature (\(^{\circ}\)C): 1350.0
  Sintering Time (hours): 6.0
  Calcination Temperature (\(^{\circ}\)C): 1200.0
  Calcination Time (hours): 4.0

- Reaction: 1 Li2CO3 + 1 MnCO3 + 0.5 O2 == 1 Li2MnO3 + 2 CO2
  Sintering Temperature (\(^{\circ}\)C): 750.0
  Sintering Time (hours): 24.0
  Calcination Temperature (\(^{\circ}\)C): 700.0
  Calcination Time (hours): 20.0

- Reaction: 0.5 Fe2O3 + 0.5 Tb2O3 == 1 TbFeO3
  Sintering Temperature (\(^{\circ}\)C): 1300.0
  Sintering Time (hours): 12.0
  Calcination Temperature (\(^{\circ}\)C): 1200.0
  Calcination Time (hours): 12.0

- Reaction: 3 BaCO3 + 1.18 CaO + 0.91 Nb2O5 + 0.135 O2 == 1 Ba3Ca1.18Nb1.82O9 + 3 CO2
  Sintering Temperature (\(^{\circ}\)C): 1600.0
  Sintering Time (hours): 2.0
  Calcination Temperature (\(^{\circ}\)C): 1400.0
  Calcination Time (hours): 4.0

- Reaction: 1 CaCO3 + 3 CuO + 4 TiO2 == 1 CaCu3Ti4O12 + 1 CO2
  Sintering Temperature (\(^{\circ}\)C): 1080.0
  Sintering Time (hours): 10.0
  Calcination Temperature (\(^{\circ}\)C): 950.0
  Calcination Time (hours): 15.0

- Reaction: 1 CaCO3 + 1 ZrO2 == 1 CaZrO3 + 1 CO2
  Sintering Temperature (\(^{\circ}\)C): 1449.85
  Sintering Time (hours): 50.0
  Calcination Temperature (\(^{\circ}\)C): 1349.85
  Calcination Time (hours): 10.0

- Reaction: 0.9 CeO2 + 0.025 O2 + 0.05 Re2O3 == 1 Re0.1Ce0.9O2
  Sintering Temperature (\(^{\circ}\)C): 1650.0
 \end{lstlisting}

\end{document}